\documentclass[fleqn,usenatbib]{mnras}

\usepackage[T1]{fontenc}
\usepackage[lofdepth, lotdepth]{subfig}

\DeclareRobustCommand{\VAN}[3]{#2}
\let\VANthebibliography\thebibliography
\def\thebibliography{\DeclareRobustCommand{\VAN}[3]{##3}\VANthebibliography}

\usepackage{graphicx}	
\usepackage{amsmath}	
\usepackage{amssymb}	
\usepackage{xcolor}

\title[]{Observations and detectability of young Suns' flaring and CME activity in optical spectra}

\author[M. Leitzinger et al.]{
M. Leitzinger,$^{1}$\thanks{E-mail: martin.leitzinger@uni-graz.at}
P. Odert,$^{1}$
R. Greimel$^{2}$
\\
$^{1}$Institute of Physics, Department for Astrophysics and Geophysics, University of Graz, Universit\"atsplatz 5, 8010 Graz, Austria\\
$^{1}$RG Science, Schanzelgasse 17, 8010 Graz, Austria}
\date{Accepted XXX. Received YYY; in original form ZZZ}

\pubyear{2015}

\begin{document}
\label{firstpage}
\pagerange{\pageref{firstpage}--\pageref{lastpage}}
\maketitle

\begin{abstract}

The Sun's history is still a subject of interest to modern astrophysics. Observationally constrained CME rates of young solar analogues are still lacking, as those require dedicated monitoring. We present medium resolution optical spectroscopic monitoring of a small sample of bright and prominent solar analogues over a period of three years using the 0.5m telescope at observatory Lustb\"uhel Graz (OLG) of the University of Graz, Austria. The aim is the detection of flares and CMEs from those spectra. In more than 1700 hours of spectroscopic monitoring we found signatures of four flares and one filament eruption on EK~Dra which has been reported in previous literature, but we complementarily extended the data to cover the latter phase. The other stars did not reveal detectable signatures of activity. For these non-detections we derive upper limits of occurrence rates of very massive CMEs, which are detectable with our observational setup, ranging from 0.1 to 2.2~day$^{-1}$, but these may be even smaller than the given rates considering observational biases. Furthermore, we investigate the detectability of flares/CMEs in OLG spectra by utilizing solar 2D H$\alpha$ spectra from Mees solar observatory. We find that solar-sized events are not detectable within our observations. By scaling up the size of the solar event, we show that with a fractional active region area of 18\% in residual spectra and 24\% in equivalent width time series derived from the same residuals that solar events are detectable if they had hypothetically occurred on HN~Peg.

\end{abstract}




\section{Introduction}

The Sun has a 4.6~Gyr long history which was subject to numerous investigations. The ``Sun in time'' program \citep[see e.g.][]{DorrenGuinan1994, Guedel2007} was founded to investigate the Sun's history in great detail. The radiation environment was reconstructed from X-rays \citep[e.g.][]{Dorren1995, Guedel1997, Telleschi2005, Guinan2017}, EUV \citep[e.g.][]{Guedel1997b, Tu2015}, FUV \citep[e.g.][]{Guinan2003}, UV \citep[e.g.][]{DorrenGuinan1994, Dalton2019}, optical \citep[e.g.][]{Messina2002} to radio \citep[e.g.][]{Guedel1994, Villadsen2014, Fichtinger2017}. The spectral energy distributions of the Sun in time has been inferred \citep{Ribas2005, Claire2012} and also the solar wind in time has been investigated \citep[][]{OFionnagain2018}.\\
Every study focusing on solar analogues of different age may be attributed to the idea of the ``Sun in time'' program.  Transient activity phenomena like flares and CMEs of the young Sun can be characterized with a significant observational effort only as those are detectable via time series observations which require much observing time. Here, especially the CME environment of the young Sun remains still relatively unknown. However, flare frequency distributions, as well as flare power laws depending on the stars X-ray luminosity of young solar analogue stars and others have been presented by \citet{Audard2000}. Based on these power laws \citet{Odert2017} have deduced relations to estimate stellar CME occurrence rates. Prior to \citet{Odert2017}, \citet{Aarnio2012} established a methodology to relate solar flare/CME relations with stellar flaring relations to infer stellar CMEs and their parameters. \citet{Drake2013} applied a similar approach and identified the problem of the unknown stellar flare-CME association rate, as extrapolating to higher energies while using solar relations leads to unrealistic high energy requirements which have been not observed yet. \citet{Osten2015} assumed energy partition between bolometric flare radiation and kinetic energy of the associated CME. These authors found mass loss rates comparable to previous studies. To explain energy budget problem discussed in \citet{Drake2013},  \citet{Odert2017} suggested then that probably the whole flare-CME association rate may shift to larger energies.\\ 
Stellar flares are a subject of ongoing research going back to the first half of the last century where stellar flares have been detected using ground-based observations \citep[][followed by numerous studies]{Joy1949, Luyten1949}. With satellite missions such as the Microvariability and Oscillations of Stars Telescope (MOST), Convection, Rotation and planetary Transits (CoRoT), Kepler and now with the Transiting Exoplanet Survey Satellite (TESS) and in the near future also with the PLAnetary Transits and Oscillation of stars mission (PLATO) long-term photometric measurements were and will be accessible. This enabled statistical investigations of flares \citep[see e.g.][]{Balona2015, Davenport2016} and superflares \citep[(E$>$10$^{33}$~erg), see e.g.][]{Maehara2012, Tu2020, Doyle2020, Tu2021, Okamoto2021}. Flare-frequency distributions from TESS or Kepler are determined for more energetic flares, as usually such broadband photometric observations are insensitive to low energetic flares, as those simply leave no signature in a light-curve and are hidden in the noise.\\
Stellar CMEs have been detected so far mainly on dMe stars \citep[e.g.][]{Houdebine1990, Guenther1997, Vida2016} using the method of Doppler shifted emission. This method uses the signature of plasma being ejected from a star. The signature, either appearing in absorption or emission is Doppler shifted by its projected velocity. This signature is often very pronounced in Balmer lines similarly to erupting filaments/prominences on the Sun. Optical spectroscopic monitoring programs to search for stellar CMEs using the method of Doppler shifted emission/absorption are often focused on dMe stars, as those are known to frequently flare and therefore also possibly may host CMEs. dG stars reveal also frequent flares if their X-ray luminosity is large, i.e. their activity level is high. However, for the detection of flares and CMEs, only stronger events, compared to dMe stars, may be detected, because of the higher continuum around H$\alpha$ on those stars, or in other words a higher contrast is favorable for stellar CME detection using the method of Doppler-shifted emission/absorption in the H$\alpha$ line.\\
Only recently \citet{Namekata2021} presented the detection of a blue wing absorption in H$\alpha$ during a superflare on the young solar analogue EK Dra, which the authors interpreted as an erupting filament. This event was simultaneously observed by two telescopes. \citet{Inoue2023} report on the detection of a high velocity blue-wing emission feature being interpreted by the authors as prominence eruption on the RS CVn system 1355~Ori, consisting of a K2-4 sub-giant and a G1 dwarf. Even more recently \citet{Namekata2024} focus again on EK~Dra and present this time two prominence eruptions from which one has a projected bulk velocity being above the escape velocity of EK~Dra and reveals a simultaneously observed candidate of coronal dimming.\\
Using the method of Doppler shifted emission/absorption numerous candidate events have been found especially on dMe stars \citep[e.g.][]{Fuhrmeister2018,Vida2019}. With this method only events with a projected bulk velocity being greater the stars' escape velocity can be treated as eruptive events. It can be concluded that they are escaping from the star, as their true velocities can be even higher. But in numerous studies events with much smaller projected bulk velocities have been found which are much more difficult to interpret, as those may also originate from flaring plasma motions. To better interpret these spectral signatures investigating the Sun seen as a star may help. Instruments doing solar 2D spectroscopy are rare. In the 90ies of the last century Mees CCD provided H$\alpha$ 2D spectroscopy of a cut-out of the solar-disk, in 2016 SMART/SDDI went into operation, doing full-disk multi-filter measurements resulting in a full-disk H$\alpha$ profile, and only recently the Chinese H$\alpha$ Solar Explorer (CHASE) providing full-disk H$\alpha$ spectroscopy. Mees did H$\alpha$ 2D H$\alpha$ spectroscopy of solar cycles 22 and 23, so representing the past, whereas SMART/SDDI operated during  the second half of cycle 24 and cycle 25 and CHASE beginning of cycle 25 up to now and hopefully also in the future. Spectroscopic Sun-as-a-star observations date back to the seventies of the last century \citep[e.g.][]{Livingston1981}. Spatial integrated investigations of flares and erupting filaments have been presented by e.g. \citet{Den1993, Ding2003, Ichimoto2017} and only recently progress has been made especially to understand Balmer line asymmetries related to erupting filaments/prominences \citep{Namekata2021,Leitzinger2021,Namekata2022b, Otsu2022, Otsu2024,Ma2024}, as the Sun is the only star where we can actually see if a CME occurred in spatial and temporal vicinity to a filament/prominence eruption.
\\
Also solar CME signatures are used to be searched on other stars, such as radio type II and IV bursts. Solar type II bursts are signatures of shock waves which can be driven by CMEs. On the shock front electrons are accelerated and radiation is emitted in dependence of the plasma frequency. Moving type IV bursts are related to trapped electrons inside CME structures originating from flare sites \citep[see e.g.][]{Gopalswamy2016}. Several attempts have been made to detect those radio signatures, at low frequencies \citep[e.g.][, in the MHz regime]{Leitzinger2009,Boiko2012,Konovalenko2012,Crosley2016} and also at higher frequencies \citep[e.g.][ in the GHz regime]{Crosley2018a, Villadsen2019, Zic2020, Bloot2024}. In nearly all of those studies bursts have been detected but so far no study claimed that a stellar analogue of a solar radio type II burst was detected. An analogue of a solar type IV burst occurring on Proxima Centauri has been presented by \citet{Zic2020}. Only recently \citet{Mohan2024} reported on the detection of a solar-like radio type IV burst on the young and nearby dMe star AD~Leo using 550-850~MHz observations from the upgraded Giant Meterwave Radio Telescope (uGMRT). \citet{Mullan2019} argued for radio quiet CMEs as the strong magnetic fields in dM stars would require very large CME velocities to produce type II emission. \citet{OFionnagain2022} investigate $\epsilon$~Eri theoretically and find that type II bursts are visible above the ionospheric cut-off frequency sufficiently long (about half an hour) if observed by LOFAR. \citet{AlvoradoGomez2020} investigate type II burst frequencies on Proxima Centauri and find that type IIs may emit below the ionospheric cut-off frequency making it unaccessible to ground-based radio facilities.\\
Continuous absorptions during flares at X-ray wavelengths have been observed and interpreted being caused by plasma clouds obscuring the stellar flaring region. \citet{Favata1999} find during a flare on Algol an increased hydrogen column density which decreased in time. This behaviour was interpreted by the authors as cool, absorbing material in the line of sight, reminiscent of an erupting filament \citep[see][for a review on this method]{Moschou2019}.\\    
Another solar CME signature is coronal dimmings \citep[for a review on solar CME detections see][]{Webb2012} which is the sudden evacuation of coronal plasma on the Sun being closely correlated to CMEs \citep[e.g.][]{Dissauer2019}. By establishing the Sun-as-a-star signature of coronal dimmings on the Sun, this method was applied to stars in \citet{Veronig2021} where around 20 potential CME candidates have been detected on late-type main-sequence stars.\\
There are also other signatures which authors related to be caused by CMEs, such as e.g. \citet{Bond2001}, who found transient absorption features in a pre-cataclysmic binary system containing a white dwarf and a K-dwarf which the authors interpreted as CMEs from the K-dwarf crossing the line-of-sight in front of the white dwarf. \citet{Senavci2018} investigate the RS CVn eclipsing binary SV Cam (F9V+K4V) and identify excess absorption which the authors interpret as cool plasma (filaments/prominences) obscuring the primary component.\\
For reviews on stellar CMEs see \citet{Moschou2019}, \citet{Leitzinger2022c}, \citet{Osten2023}, \citet{Tian2023}.\\
For the detection of stellar CMEs the approaches are manifold as can already be seen from the methodologies mentioned above. For the method of Doppler shifted emission/absorption the observing strategies are manifold as well. One can either observe single stars or one uses multi-object spectroscopic devices which are able to observe several stars at the same time spectroscopically \citep[see e.g.][]{Guenther1997, Leitzinger2014, Korhonen2017}. Both approaches have their advantages but also their drawbacks. Single star observations are of course very time consuming, because when one focuses not only on one star then observing time increases by factors in the order of the number of stars one wants to observe. For multi-object observations one catches in one exposure several stars but the target selection is of course limited. Existing multi-object spectrographs have very differently sized fields-of-views (FOVs), from square arcmin to square degrees. But this means always one has to select open clusters or associations to fill the FoV of the instrument with a sufficiently large number of targets of similar age. Moreover, multi-object spectrographs are mainly available on large telescopes where usually the access is restricted to member institutions or telescope time is distributed via a competition process. Experience has shown that observing proposals with the aim of monitoring stars are not likely granted with observing time as such proposals would block a telescope over a long period. Therefore another approach needs to be chosen for monitoring of solar-analogues. At the authors home institution a small-sized telescope is available. The 0.5m telescope is equipped with a slit spectrograph and a back-end device which was optimized for this purpose. The minimum requested signal-to-noise (S/N) for the detection of CMEs within 10-15 minutes exposures, to properly resolve CME evolution, is $\sim$20. This restricts the target selection to brighter stars ($<$7~mag). There is a small number of young solar analogues which fulfill this criterion (see section~\ref{targetstars}). As the telescope time can be attributed up to 75\% of the total observing time we spectroscopically monitored solar analogues for three years to search for optical signatures of flares and erupting filaments/prominences in H$\alpha$ using the method of Doppler-shifted emission/absorption. .
\section{Observations, data, and targets}
\subsection{Observatory Lustb\"uhel Graz (OLG)}
The OLG is situated at the border of the city of Graz, Austria. It contains an astronomy part belonging to the Institute of Physics, Department for Astrophysics and Geophysics of the University of Graz, a satellite geodesy part, belonging to the Space Research Institute of the Austrian Academy of Sciences, and a telecommunications part belonging to the University of Technology, Graz, Austria. The building hosts two domes, one belonging to the astronomy part, hosting a 0.5m reflector manufactured by the Astro Systeme Austria (ASA) company and the other to the satellite geodesy part. The ASA 0.5m is a german mount telescope with a focal length of 4.5m. Separated from the building exists a tower with a third dome hosting the Ballistische Mess Kammer (BMK), a rare astrograph manufactured by ZEISS from the nineteen seventies, with a focal length of 0.75m yielding an immensely large angular FoV of 19.3 degrees.\\
\subsubsection{Instruments}
The ASA 0.5m telescope is equipped with four instruments mounted on a custom manufactured so-called multi-port, a device mounted in Cassegrain focus with a rotating mirror feeding light to the instruments. Mounted to the multi-port are two imaging cameras with broad and narrow-band filters, one video camera for asteroid occultation measurements, and a spectrograph with a dedicated CCD camera.\\
The spectrograph is a Littrow High Resolution Spectrograph (LHIRES), an off-the-shelf slit spectrograph with adjustable wavelength range and removable gratings. The spectrograph back-end is an Apogee Alta F47 back-illuminated CCD camera reaching a quantum efficiency of $>$90\% around H$\alpha$. For spectroscopic monitoring we select the medium/low resolution grism with 600 groves/mm corresponding to a spectral resolving power of R=2700 which corresponds to $\sim$100~km~s$^{-1}$ at H$\alpha$. This is a resolution comparable to the now already decommissioned VIsible Multi Object Spectrograph (VIMOS) at the Very Large Telescope (VLT) of the European Southern Observatory (ESO) with its orange grism which we have used in the past \citep[see][]{Leitzinger2014}. The used grism yields a spectral range of $\sim$5800 .. 6800\AA{}.
\begin{table*}
	\centering%
	\caption{Target stars of the study, their characteristics \citep[taken from][]{Guedel2007}, number of observed spectra, and flare rates. The XUV flare rates have been estimated from the X-ray luminosities using the power-law from \citet{Audard2000}, the cumulative H$\alpha$ flare rates have been estimated from the XUV flare rates based on \citet{Leitzinger2020}, and the TESS flare rates have been determined by eye from the light curves. HN~Peg and $\kappa^{1}$~Cet did not show any flares in their TESS light curves therefore we give here upper limits only.}%
	\label{peakfluxtable}
	\begin{tabular}{lcccccccc} 
  name             &    age   &   log~Lx     &   V   & P$_{\mathrm{rot}}$ & number of spectra &      XUV flare rate         &     H$\alpha$ flare rate    & TESS flare rate \\
                   &   [Gyr]  & erg~s$^{-1}$ & [mag] &        day         &                   & N(>10$^{32}$erg) day$^{-1}$ & N(>10$^{32}$erg) day$^{-1}$ &    day$^{-1}$   \\
\hline
 EK Dra            &    0.1   &    29.93     &  7.6  &        2.75        &       6722        &             52              &              2.81           &       0.54      \\
 HN~Peg            &    0.3   &    29.12     &  5.9  &        4.86        &       13433       &             9               &              0.48           &    $<$0.11      \\  
$\pi^{1}$~UMa      &    0.3   &    29.10     &  5.6  &        4.68        &       5112        &             8               &              0.46           &       0.06      \\ 
$\chi^{1}$~Ori     &    0.3   &    28.99     &  4.4  &        5.08        &       6856        &             7               &              0.36           &       0.04      \\
$\kappa^{1}$~Cet   &    0.75  &    28.79     &  4.9  &        9.2         &       648         &             4               &              0.23           &    $<$0.06      \\
 \hline
	\end{tabular}
\end{table*}
\subsubsection{Observing}
Observing at OLG can be done almost automatically. Co-author R. Greimel developed a python based software with user interface which combines all the necessary software packages/tools needed to operate the telescope, dome, and instruments. For the spectrograph a tool was developed which tracks the star in the image of the guide camera hereby keeping the spectrograph slit on the target star. This enables that the observer needs only to start the system, take calibration frames, select the target of the night, and start observing. The software system also includes an auto-close option which closes the dome, parks the telescope, and disconnects all the software tools from the python based software. To further ease the observer's work the whole system can be operated remotely and therefore the observer does not need to be at the observatory to start the system. Furthermore OLG has a weather station including all-sky camera, wind, temperature, and humidity measurements, all of that is necessary, especially for remote observing.
\subsection{Target stars}
\label{targetstars}
As already mentioned in the introduction, according to the aperture of the telescope and the location of the observatory, we are restricted to bright solar analogues. We decided to select only a handful of solar analogues to dedicate sufficient observing time to each target to enhance the chance to detect signatures of flares and CMEs on those stars. We used the list of the ``Sun in time'' sample \citep{Guedel2007} to compile a target list. In table~\ref{peakfluxtable} we list the target stars of our study together with characteristics and the number of recorded spectra for each target star. \\
\textit{\textbf{EK~Dra:}} EK~Dra is a $\sim$100~Myr solar analogue star of spectral type dG0 with a logLx being $\sim$100 times larger than for the Sun \citep[see e.g.][]{Guedel2007, Senavci2021} and rotating 10 times faster than the Sun \citep{Koenig2005}. EK~Dra has a wide companion with a period of $\sim$45~years having roughly half the mass of the primary \citep{Koenig2005}.\\
\textit{\textbf{HN~Peg:}} HN~Peg is $\sim$300~Myr solar analogue of spectral type dG0 with a logLx being $\sim$50 times larger than the solar value \citep[see e.g.][]{Guedel2007} and rotating $\sim$6 times faster than the Sun. HN~Peg has a T-dwarf companion with a separation of $\sim$800AU \citep[][]{Luhman2007}.\\
\textit{\textbf{$\pi^{1}$~UMa:}} $\pi^{1}$~UMa has also an age of 300~Myr and a logLx being 40 times larger than that of the Sun \citep[see e.g.][]{Guedel2007} similar to HN~Peg and $\chi^{1}$~Ori. Also the rotation period is comparable to the ones of HN~Peg and $\chi^{1}$~Ori making it rotate $\sim$6 times faster than the Sun. $\pi^{1}$~UMa is a single star with no known companion.\\ 
\begin{figure}
\begin{center}
\includegraphics[width=\columnwidth]{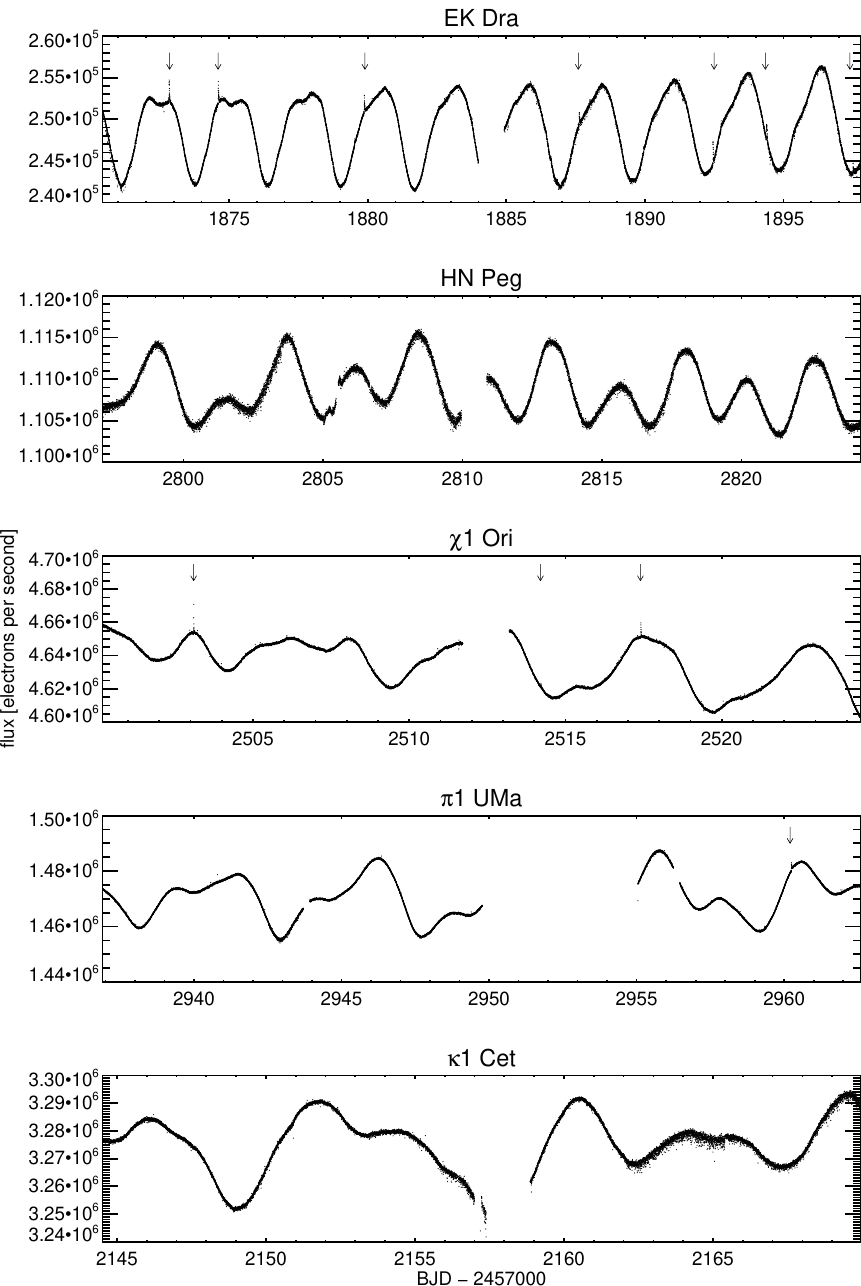}
 \caption{TESS light curves of the target stars from top to bottom for EK~Dra (sector 21), HN~Peg (sector 55), $\chi$1~Ori (sector 44), $\pi$1~UMa (sector 60), and $\kappa$1~Cet (sector 31). Arrows mark flares in the light curves.\label{TESSlcs}}

\end{center} 
\end{figure}
\textit{\textbf{$\chi^{1}$~Ori:}} $\chi^{1}$~Ori is also a $\sim$300Myr solar analogue of spectral type dG1 with a logLx being $\sim$30 times larger than that of the Sun \citep[see e.g.][]{Guedel2007} and rotating $\sim$5 times faster than the Sun, similar to HN~Peg. $\chi^{1}$~Ori has a companion of spectral type M with a period of $\sim$14~years and a mass of one seventh of the mass of the primary. \\
\textit{\textbf{$\kappa^{1}$~Cet:}} $\kappa^{1}$~Cet is the oldest star in the sample with its 750~Myr. Its spectral type is dG5 and its logLx is $\sim$20 times larger than the solar value \citep[see e.g.][]{Guedel2007}. $\kappa^{1}$~Cet spins $\sim$3 times faster than the Sun. Companions of $\kappa^{1}$~Cet have been searched but to date none has been confirmed.\\

\subsection{Other data and data reduction}
For deducing TESS flare rates we use all available TESS 2 minute light curves for the stars in our target sample (see Fig.~\ref{TESSlcs} for examples). For the identification of flares we use the already extracted light-curves available from TESS. For EK~Dra we use sectors 14, 15, 16, 21, 22, 23, 41, 48, 49, 50, for HN~Peg sector 55, for $\chi^{1}$~Ori sectors 43, 44, 45, for $\pi^{1}$~UMa sectors 20, 47, 60, and for $\kappa^{1}$~Cet sectors 4, 31. As the number of light curves to be searched for flares is small (19) we analysed the light curves by eye. EK~Dra shows in every TESS light curve flares. $\pi^{1}$~UMa reveals only one flare, in one out of three TESS sectors. $\chi^{1}$~Ori reveals only three flares in one out of three TESS sectors and HN~Peg and $\kappa^{1}$~Cet reveal no flares in their TESS light curves.\\
For a comparative analysis of solar activity phenomena (see section~\ref{solobsres}) we utilize data from Mees CCD imaging spectrograph \citep[MCCD,][]{Penn1991} at Mees Solar Observatory (MSO). These data are two dimensional imaging spectroscopic data and are available as level 0.5 and/or level 1.0 data. Level 0.5 data are raw frames with a fits header whereas level 1.0 data are nominally processed. For the solar analysis we used level 1.0 data with a modified implementation of the default flat-fielding code (priv. comm. KD~Leka).\\
Optical spectroscopic observations for the study started in June 2018 and lasted until September 2021, with a break between November 2018 and March 2019 where the camera had to be sent back for electronic adjustment. For spectroscopic observations we used only half of the chip to minimize read-out time and save hard drive memory. The thereby reduced number of pixels was still sufficient to determine the background properly. The usual exposure time was set to three minutes as this ensures a proper temporal resolution and in case of low signal, exposures can be added without increasing noise significantly. The data were reduced with the Imaging and Reduction Analysis Facility (IRAF) using the package $specred$ and background was subtracted using the task $apscatter$. The spectra were wavelength calibrated and normalized before analysis. The spectra show a drift per night due to the fact that the spectrograph is Cassegrain mounted and therefore flexure is an issue. The drift in wavelength needs to be removed which we have done using cross correlation of the actual spectra with a reference spectrum per night. This yielded one dimensional spectra which were then further processed using the Interactive Data Language (IDL). In total we recorded 32771 spectra for the investigation of activity of solar analogues (see table~\ref{peakfluxtable}).

\begin{figure}
\begin{center}
\includegraphics[width=\columnwidth]{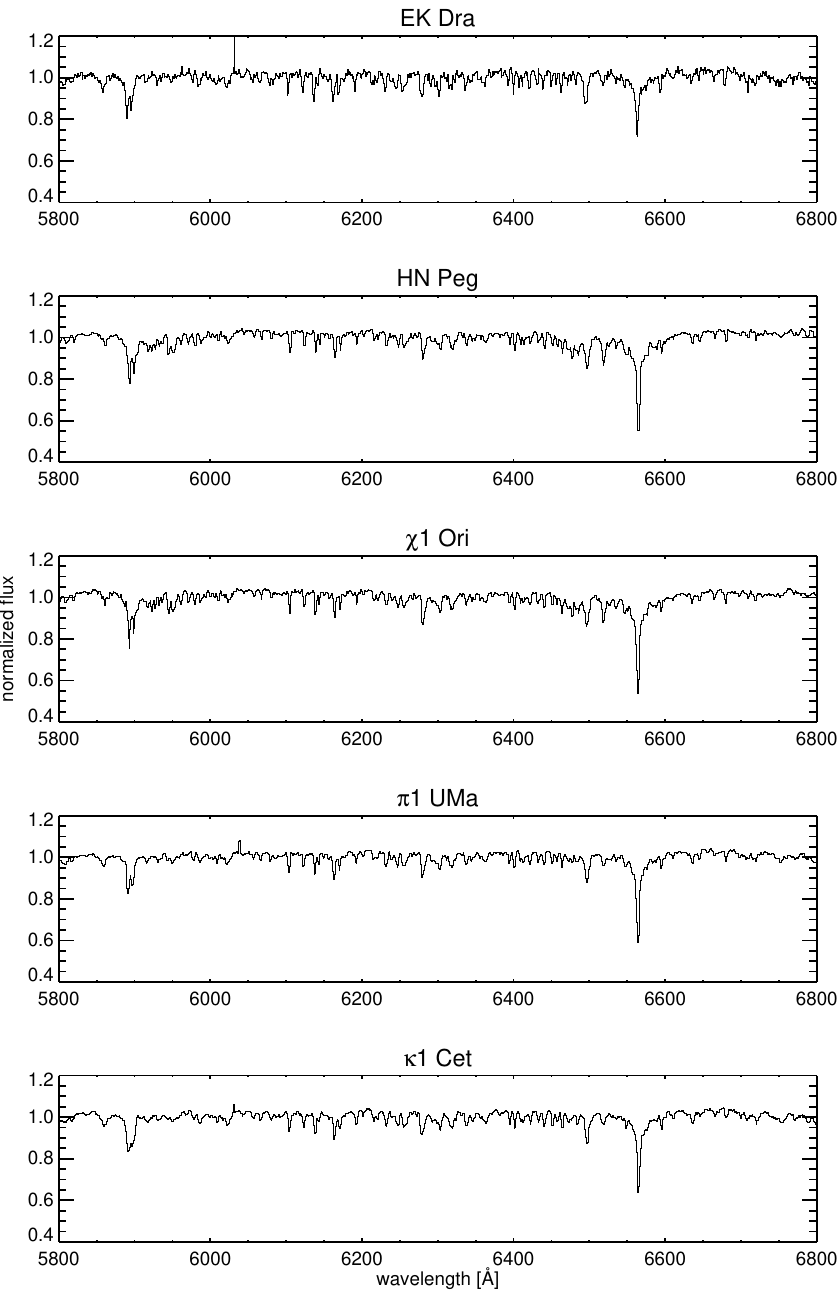}
 \caption{Examples of OLG spectra of the target stars from top to bottom for EK~Dra, HN~Peg, $\chi$1~Ori, $\pi$1 UMa, and $\kappa$1~Cet. \label{OLGspec}}
\end{center} 
\end{figure}



\section{Results}
\subsection{Stellar observations}
We analyzed all OLG spectra in terms of variability in H$\alpha$ by computing the residual spectra by subtracting a quiescent spectrum from the time series. The quiescent spectrum is represented by the mean of all spectra from one time series, i.e. one night, after rejecting noisy spectra. In these residuals we did not see signatures of flares or CMEs. 
As a second step, we then binned the spectra in time up to 30 minutes to increase S/N.\\
The typical S/N of the stellar spectral sample vary from star to star. For the nominal integration times (3~min) we derive a S/N distribution for EK~Dra peaking at a value of 60, for HN~Peg peaking at a value of 80, for $\pi^{1}$UMa peaking at the same value, for $\chi^{1}$Ori peaking at a value of 110, and for $\kappa^{1}$~Cet peaking at a value of 90. The S/N values behave accordingly with the stars magnitudes, which is reasonable. The S/N values were calculated with the algorithm presented in \citet{Stoehr2008} which represents a general way to derive S/N from one-dimensional spectra. Temporal binning of up to 10 spectra yields S/N distributions with peak values being 30-40\% larger than that derived from the single spectra, but with  maximum values being a factor 2 larger than the S/N distributions derived from the single spectra. The algorithm from \citet{Stoehr2008} seems to be more conservative in deriving S/N values when compared to the rough estimation of spectral noise for bright stars being calculated as the square root of the stellar flux.\\  
Of course, with temporal binning the number of temporally binned spectra decreases, so that we end up with 20 spectra per night at maximum, corresponding to one full winter night of $\sim$10 hours. Also here we produced residual spectra. We produced a catalogue of spectra and residual spectra which we then analysed by eye. Even if the amount of nights is quite large, signatures lying above the noise are easily visible, and therefore a search for signatures is feasible by eye. This revealed signatures which were invisible in the original unbinned three minute exposures. We did not see any signatures of flares and CMEs in the target stars except for EK~Dra being the
\begin{figure}
\begin{center}
\includegraphics[width=\columnwidth]{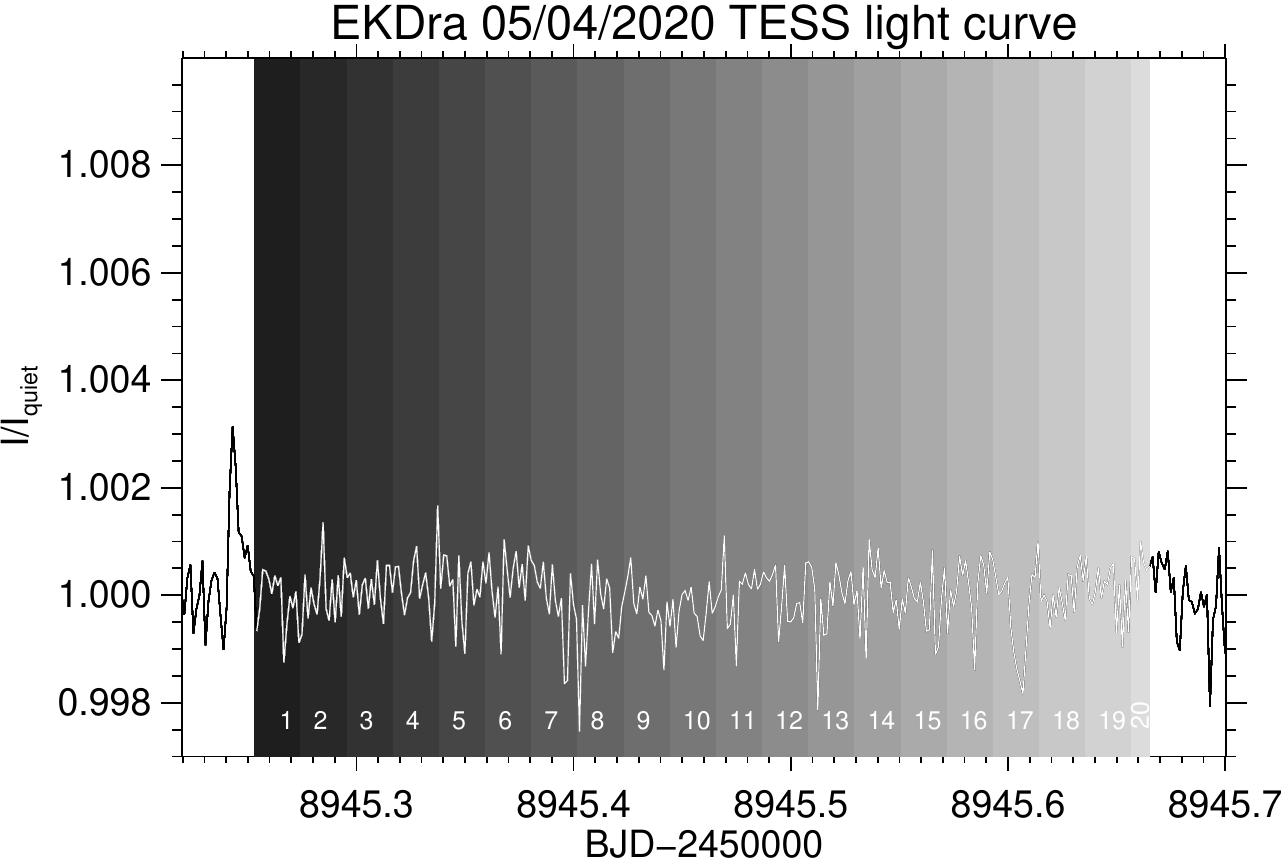}
\includegraphics[width=\columnwidth]{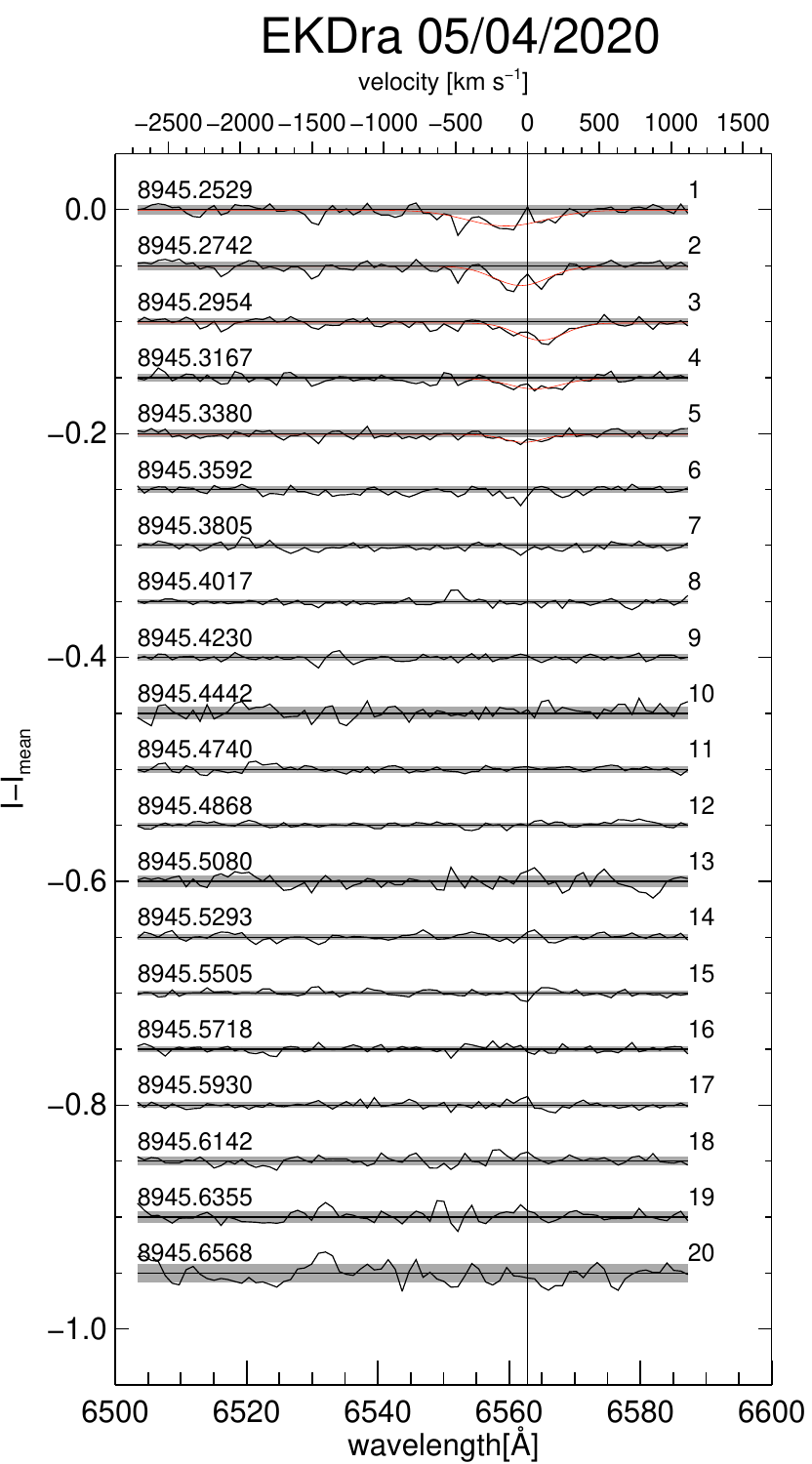}
\caption{Upper panel: TESS lightcurve coinciding in time with the residual spectral time series shown in the lower panel. Differently grey-shaded areas denote the residual spectra in the lower panel. The indexing corresponds to the indexing in the lower panel. Lower panel: residual spectral time series (from top to bottom) of EK~Dra from the night of the 5th April 2020. Grey shaded areas denote the 1-$\sigma$ uncertainty of the residuals. Red solid lines are gaussian fits to the absorption visible from residuals 1-5.\label{olgevents}}
\end{center} 
\end{figure}
\begin{table*}
	\centering%
	\caption{Detected flares/CMEs and their characteristics from the present study (all from EK~Dra). Calculation of the flare energies is described in the text.}%
	\label{resulttable}
%
%
	\begin{tabular}{lcccccc} 
 observing night  & flare & CME  & starting time H$\alpha$/TESS &  flare duration H$\alpha$/TESS  &      flare energy H$\alpha$/TESS          &   H$\alpha$ asymmetry    \\
                  &       &      &              [BJD-2450000]   &              [min]              &                    [erg]                  &                          \\
\hline
 19/02/2020       &  yes  & no   &     8899.5174/8899.5322      &           150/54                & 2.3$\times$10$^{32}$/3.5$\times$10$^{33}$ &           red            \\
 04/04/2020       &  yes  & no   &     8944.5918/8944.5791      &            60/48                & 6.8$\times$10$^{31}$/7.0$\times$10$^{33}$ &           red            \\
 05/04/2020$^{1}$ &  no   & yes  &     8945.2529/-              &             -/-                 &                    -/-                    &           blue           \\  
 14/04/2020       &  yes  & no   &     8954.2819/8954.3164      &           240/60                & 8.0$\times$10$^{32}$/1.7$\times$10$^{33}$ &           red            \\ 
 22/04/2020       &  yes  & no   &     8962.5232/-              &           180/-                 & 6.9$\times$10$^{32}$/-                    &           red            \\
%
%
 \hline
	\end{tabular}
\footnotesize
\begin{flushleft}
(1)We know from the TESS light curve and \citet{Namekata2021} that there was a flare prior to event 05/04/2020, but we did not detect the flare by the spectroscopic monitoring, as the spectroscopic time series that night started after the flare.
\end{flushleft}
\end{table*}
\normalsize
youngest and most active but also the faintest solar analogue in the sample, and therefore the star with the lowest S/N in the spectra.\\
We observed EK~Dra for $\sim$336~hours which is a similar amount of time as for $\chi^{1}$Ori or for $\pi^{1}$UMa. TESS observed EK~Dra in sectors 14-16, 21-23, 41, and 48-50, and we observed coordinated with TESS in sectors 21-23. The TESS light curves of sectors 21-23 are shown in the appendix in Fig.~\ref{EKDraTessOLG}. We have indicated the times where we have observed spectroscopically by blue shaded areas. What becomes obvious is that we missed all larger flares. However, we caught some of the smaller flares, namely three in TESS sector 23 named after the nights in which they occurred, event 19/02/2020  (see left panel of Fig.~\ref{olgevents2}), event 04/04/2020 and 14/04/2020 (see Fig.~\ref{olgevents1}). Another flare was detected after TESS sector 23, so not coordinated with TESS, named event 22/04/2020 (see right panel of Fig.~\ref{olgevents2}). An absorption signature, reminiscent to a solar eruptive filament, named event 05/04/2020 in TESS sector 23 was seen as well. All of the TESS flares were superflares which can be seen from the estimated TESS flare energies given in Table~\ref{resulttable}. In Table~\ref{resulttable} we list starting times of the detected flares and the eruptive filament on EK~Dra, flare durations, and flare energies derived from TESS and/or H$\alpha$ data.
The TESS flare durations were calculated according to the pre-flare level in the light curves and the time when the light curves reach pre-flare levels again. The TESS flare energies are calculated by extracting the TESS flux (8.12$\times$10$^{-9}$erg/s/cm$^{2}$/\AA{}) from the TESS magnitude of EK~Dra \citep[Tmag=7.04,][]{Paegert2022}, using the distance to EK~Dra \citep[d=34.45~pc,][]{Petit2014} and then integrating the residual light curves (in flux units) over the flare durations. We also list the H$\alpha$ flare energies which we derive from the residual spectral time series by applying the 
quiet continuum flux around H$\alpha$ derived from GAIA DR3 spectra of EK~Dra (2.75$\times$10$^{-12}$erg/s/cm$^{2}$/\AA{}) which are available as flux calibrated spectra \citep{deAngeli2023}. By involving the distance to EK~Dra and integrating over the H$\alpha$ flare durations we are able to obtain H$\alpha$ flare energies. These energies are lower limits as we do not take into account the continuum increases during the H$\alpha$ flares However, for our flares which have been observed by TESS simultaneously, we know that the continuum variation is small (see description of single events below) and therefore negligible. This does not account for flare 22/04/2020 as this one was not observed simultaneously by TESS.\\
In the following we describe the events from Table~\ref{resulttable} in more detail.\\
\textbf{\textbf{Event in night 19/02/2020} in TESS sector 22:} TESS captured a relatively small flare with an amplitude of $\sim$0.3\% above background which lasted for about an hour. The impulsive flare phase started around BJD(-2450000) 8899.5174 and after $\sim$5~min the flare had reached its peak. The spectroscopic observations started that night one binned spectrum prior to the flare as detected by TESS (BJD(-2450000)~8899.4983, see Fig.~\ref{olgevents2}). Already the second spectrum covers the impulsive TESS flare phase and spectra 3 and 4 cover the decay phase of the TESS flare (see TESS light curve, upper panel of Fig.~\ref{olgevents2}). Taking a look at the residual spectra in the left lower panel of Fig.~\ref{olgevents2} reveals that in H$\alpha$ the flare signature can be seen beyond the decaying phase of the TESS flare deduced from TESS broadband photometry until spectrum 6. The residual spectra show a peak being red shifted by one wavelength bin (similarly as event 14/04/2020 in TESS sector 23).\\ 
\textbf{\textbf{Event in night of 04/04/2020} in TESS sector 23:} This TESS flare starting at BJD(-2450000) 8944.575 and reaching its peak in $\sim$10~min is also a relatively weak flare with an amplitude of $\sim$0.4\%. The flare was spectroscopically captured at the end of the night. The flare peak is covered by spectrum no. 13 which does not reveal a significant increase in flux at H$\alpha$. In the successive spectrum a distinct, broad (FWHM=10.4\AA{}), and red-shifted ($\Delta\lambda$=2.1\AA) signature is visible which gains flux in the following residuum. All this happens in the decaying tail of the TESS flare. As the light curve returns to quiescent levels the red-shifted asymmetry vanishes in the last residual spectrum.\\
\textbf{\textbf{Event in night of 14/04/2020} in TESS sector 23:} This TESS flare starts at BJD(-2450000) 8954.316  and reaches its peak around 15~min later. After 30 more minutes the flare has reached quiescent levels again. The peak of the flare is only $<$0.2\% above the background. The residual spectra reveal, that there is already excess flux in the residual spectrum prior the residual spectrum covering the impulsive phase. The excess flux in the residual spectra peaks during the TESS flare decay phase (cf. right lower panel of Fig.~\ref{olgevents1}). We still see some excess emission in the H$\alpha$ residual spectra although the broadband flare has already ended. Until residual spectrum 9 (cf. right upper panel of Fig.~\ref{olgevents1}) we see excess emission and this is five residual spectra or $\sim$2.5~hr later. Also here (similarly to event 19/02/2020 in TESS sector 22) we see a red-shift of the excess emission peak by one wavelength bin.\\
\textbf{\textbf{Event in night 22/04/2020}:} This flare was not observed coordinated with TESS. From the residual spectra (BJD(-2450000) 8962.5232 .. 8962.6082; see Fig.~\ref{olgevents2}) we see that there are broad emission features (FWHM=6.8-9.7\AA{}) in H$\alpha$. These flare signatures are detected in five spectra indicating a total duration of $\sim$2.5~hr. As time evolves the signatures move to the red, shifted at maximum by 3.2\AA{} which corresponds to $\sim$150~km~s$^{-1}$.\\
\textbf{\textbf{Event in night 05/04/2020} in TESS sector 23:} TESS detected a flare starting at BJD(-2450000) 8945.238 which reached its peak after seven minutes. The decay of the flare lasts for $\sim$15~min. Our coordinated spectroscopy started at the time when the star had already returned to its quiescent level in the TESS light curve. We recognized a significant absorption being visible already in the first spectrum lasting for five spectra (see Fig~\ref{olgevents}), i.e. 2.5~hr. This broad absorption feature (FWHM=7.5 .. 14.9\AA{}) moved from the blue to the red, crossing  the H$\alpha$ core and then moving back to the H$\alpha$ core wavelength. This event is reminiscent to a solar filament eruption, for a detailed discussion see section~\ref{discussion}.\\
As we have detected no flares and/or CMEs for the remaining stars in the sample we may derive upper limits only. As in \citet{Leitzinger2020} we use the relation -ln(1-0.95)/t$_{obs}$ from \citet{Gehrels1986} to derive upper limits of CME rates for the target stars with no detections. Doing so for our observations yields an upper limit for the CME/flare rate of  HN~Peg of 0.11~day$^{-1}$ ($\sim$1 in 5 days), of $\chi^{1}$~Ori of 0.21~day$^{-1}$ ($\sim$1 in 5 days), of $\pi^{1}$~UMa of 0.28~day$^{-1}$ ($\sim$1 in 4 days), and of $\kappa^{1}$~Cet of 2.22~day$^{-1}$. 
For EK Dra we have detected only one event which resembles a filament eruption, which showed a maximum projected velocity of the bulk material being close to the stars’ escape velocity \citep[cf.][]{Namekata2021}, and therefore has likely evolved into a CME. The observed CME rate yields 0.07$^{+1.61}_{-0.058}$~day$^{-1}$, which is a lower limit as the true CME rate must be higher as what we are able to detect with our observational setup. EK~Dra is also the only star in the sample where we have detected flares in H$\alpha$, therefore the observed flare rate yields 0.29$^{+0.229}_{-0.139}$ H$\alpha$ flares~day$^{-1}$. In section~\ref{discussion} we relate those values to the upper limits of our target stars in \citet{Leitzinger2020} as well as 
\begin{figure*}
\begin{center}
\includegraphics[width=16cm]{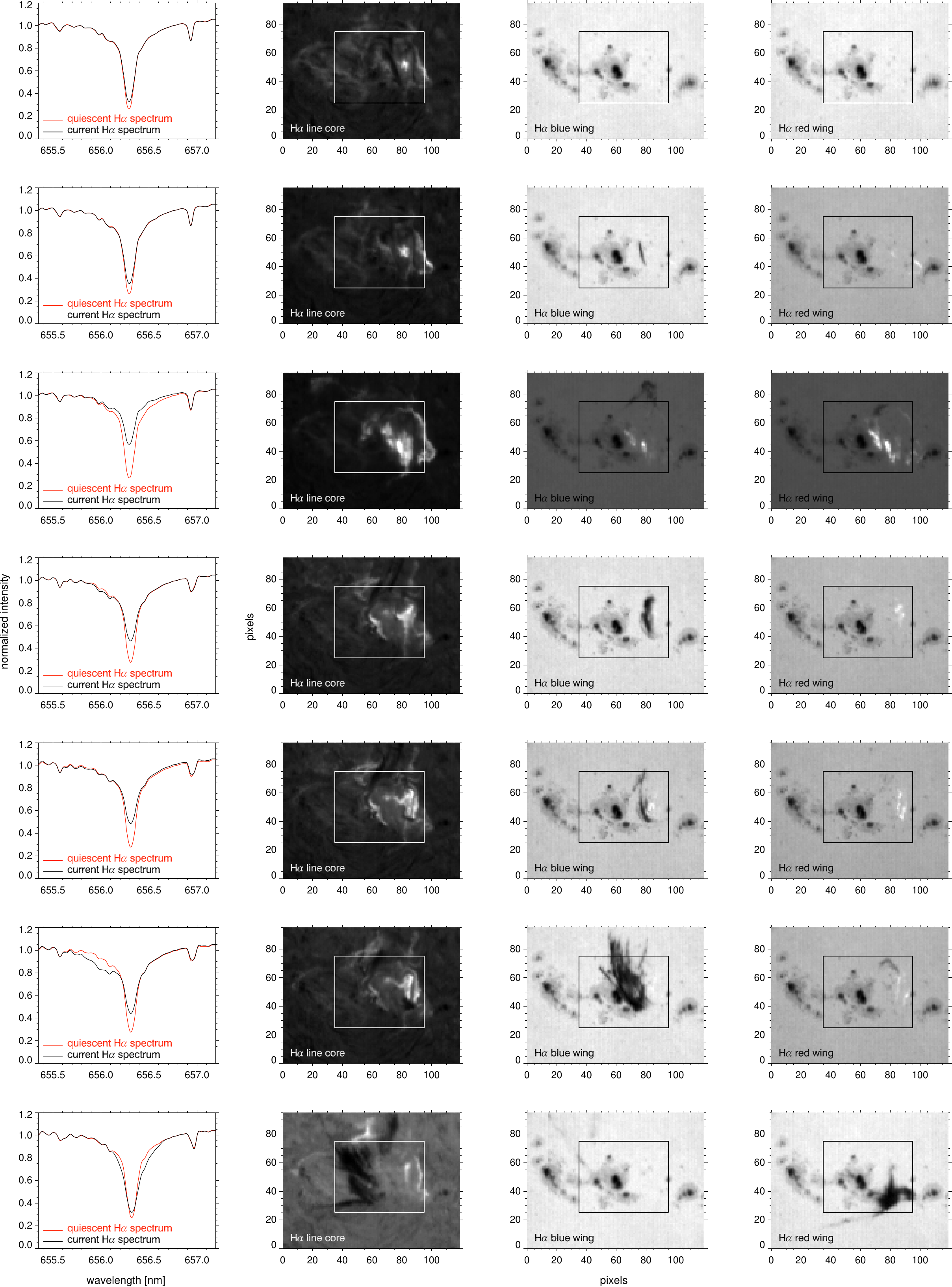}
 \caption{First column: Spatially integrated (for integration area see white/black rectangles in columns 2-4) H$\alpha$ spectra (black: current, red: quiescent). Second column: H$\alpha$ core image (at 6562.3\AA{}). Third column: H$\alpha$ blue wing image (at 6560.2\AA{}). Fourth column: H$\alpha$ red wing image (at 6565.4\AA{}). Every row (from top to bottom) corresponds to the numbering in Fig.~\ref{mccdflare} (dot symbols). \label{eventfig}}
\end{center} 
\end{figure*}
relate the upper limits to expected CME rates \citep{Odert2017, Odert2020}. We will also discuss the expected flare rates as well as the recent detection of a filament eruption on EK~Dra \citep{Namekata2021}.
\subsection{Solar observations}
\label{solobsres}
To evaluate the detectability of flares and CMEs on our target stars we construct Sun-as-a-star H$\alpha$ spectra of erupting filaments and flares (see Fig.~\ref{eventfig}), utilizing observations from MCCD at MSO which has observed the Sun between 1991 and 2007, covering thereby two maxima of solar cycles 22 and 23. In its flaring mode MCCD has a temporal resolution of $\sim$16~s, a spectral resolution of 0.375\AA{} and a FoV of 3.8$\times$4.8~arcmin.
\begin{figure}
\begin{center}
\includegraphics[width=\columnwidth]{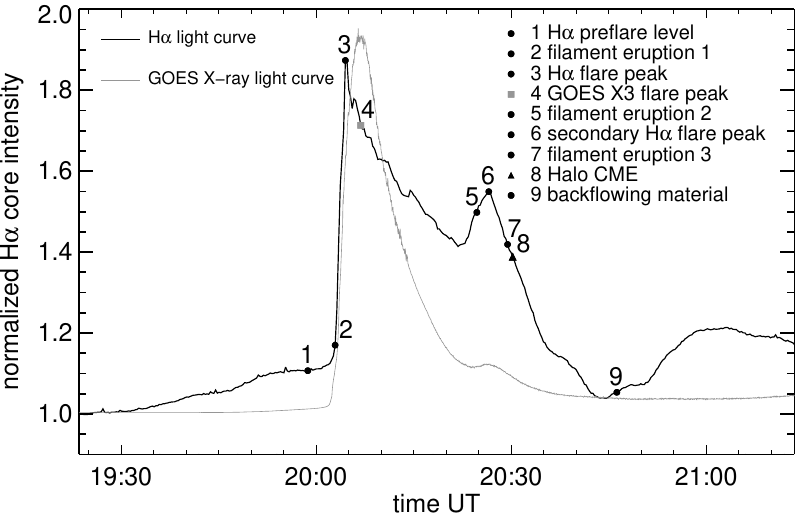}
 \caption{Solar time series of the H$\alpha$ core intensity of the 15/07/2002 event (black solid line) overplotted with the corresponding GOES X-ray light curve (grey solid line). Numbers correspond to times when flares, filament eruptions, backflowing material, and a Halo CME occurred (see text). \label{mccdflare}}
\end{center} 
\end{figure}
The data archive\footnote{\url{https://www.cora.nwra.com/MCCD/}} is freely accessible and contains numerous flares and filament/prominence eruptions. Therefore we selected all X-, M-, and C-class flares from GOES\footnote{\url{https://www.ngdc.noaa.gov/stp/space-weather/solar-data/solar-features/solar-flares/x-rays/goes/xrs/}}, and crossmatched the peak times of the flares with the observing times at MCCD/MSO. Thereby we identified $\sim$40 filament/prominence eruptions. The most significant event is the filament eruption and X-class flare from 15/07/2002, which is a well investigated event \citep[see e.g.][]{Liu2003, Gary2004, Li2005}. We selected this event to investigate the detectability of flares and CMEs on solar analogues as well as to determine the characteristics of active regions causing such eruptions. We construct Sun-as-a-star spectra by summing up all pixels belonging to the rectangle, which we have defined to include the flaring area as well as the eruptive filaments. For the strong filament eruption shown in Fig.~\ref{eventfig} in the sixth row (from the top) one can see that our defined rectangle does not cover the complete area of the filament, which does not significantly affect the spatially integrated H$\alpha$ spectrum. The 15/07/2002 event started with an X3 flare at 20:03UT on 15th July 2002 near disk center in active region NOAA 10030 followed by a filament eruption. Around half an hour later a second weaker flare peaked, occurring during the decaying tail of the X-class flare, with another filament eruption. A Halo CME was seen at 20:30UT and at 21:06 a partial Halo CME occurred.\\
In Fig.~\ref{mccdflare} we show the H$\alpha$ core intensity of the MCCD observations of the event from 15/07/2002 (black solid line) overplotted with the corresponding GOES X-ray light curve (grey solid line). We indicate seven different times in the light curve (dot symbols) for which we show the spatially integrated H$\alpha$ spectrum, together with the H$\alpha$ core, blue-(at $\sim$656.02~nm) and red-wing $\sim$656.54~nm) image in Fig.~\ref{eventfig}. In addition we show also the position of the peak of the corresponding GOES X-ray light curve on the H$\alpha$ light curve as well as the occurrence of the associated Halo CME. Number 1 in Fig.~\ref{mccdflare} corresponds to the H$\alpha$ pre-flare stage, number 2 indicates the onset of the first
\begin{figure}
\begin{center}
\includegraphics[width=\columnwidth]{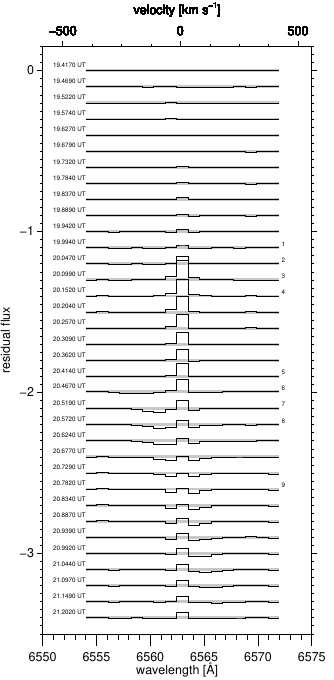}
 \caption{Time series (from top to bottom) of spectral residual flux extracted from the active region in Fig.\ref{eventfig} (white rectangle) of the solar event from 15-07-2002. The residual spectra have been scaled with a factor resulting in an area scaling of 18~\%, which is the scaling with which the spectral signatures of the eruptive filament are detectable exceeding 1-$\sigma$. The grey shaded area corresponds to the median 1-$\sigma$ uncertainties from the spectral residuals (original 3 minute spectra) of all events detected on EK~Dra (see Table.~\ref{resulttable}). Each residual spectrum is tagged with a time which corresponds to the times shown in Fig.~\ref{mccdflare}, as well as selected residual spectra with a number which corresponds to the indexing in Fig.\ref{mccdflare}.\label{solstell}}
\end{center} 
\end{figure}
\begin{figure}
\begin{center}
\includegraphics[width=\columnwidth]{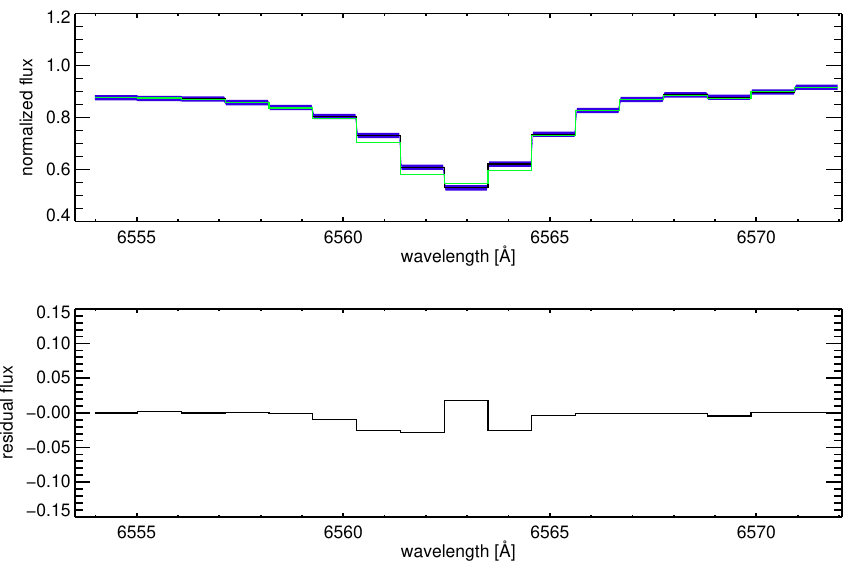}
 \caption{Upper panel: Cut-out of a normalized HN~Peg H$\alpha$ spectrum (black solid line). The blue area marks the standard deviation. The green solid line represents the spatially integrated solar MCCD spectrum of one time step (t=20.6~hour in Fig.\ref{olgmccdarea}) of the erupting solar filament from 15-07-2002 (see lower panel) superimposed on the HN~Peg spectrum. From these superimposed spectra (green solid line) we determined the EW time series shown in Fig.\ref{olgmccdarea}. Lower panel: Solar spatially integrated residual H$\alpha$ spectrum of one time step of the erupting solar filament from 15-07-2002. The applied area scaling is here 24\%. \label{olgmccdhnpeg}}
\end{center} 
\end{figure}
\begin{figure}
\begin{center}
\includegraphics[width=\columnwidth]{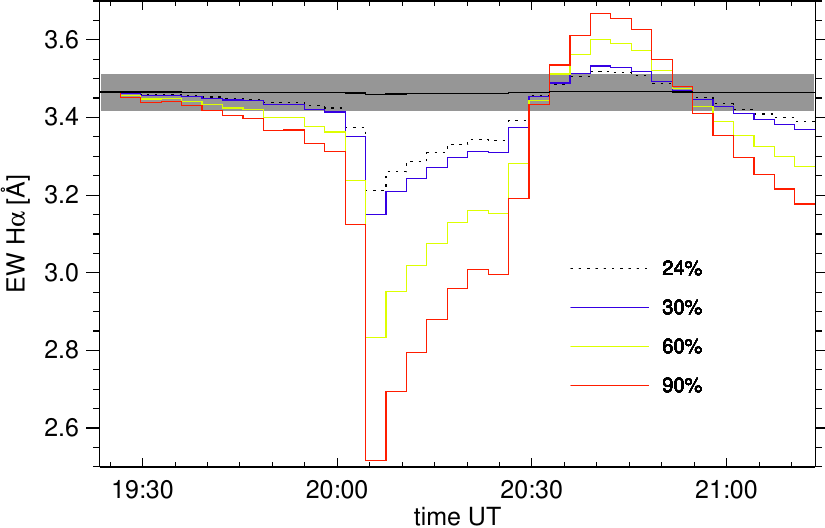}
\includegraphics[width=\columnwidth]{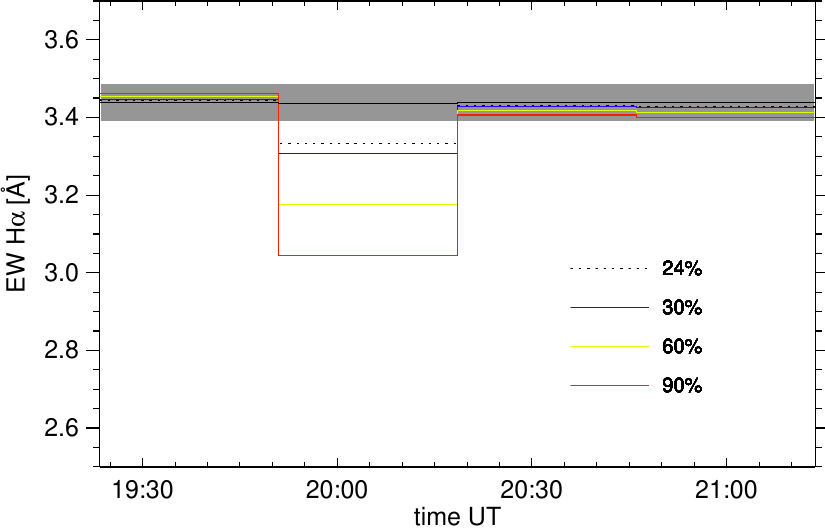}
 \caption{Upper panel: Time series of equivalent widths of the solar H$\alpha$ residual profile of the 15/07/2002 event superimposed on the  H$\alpha$ profile of HN~Peg of the original 3~min exposures. The depression starting at 20:05~UT corresponds to the X3 flare (a flare causes a decrease in EW and an increase in flux), the increase at 20:35~UT corresponds to the very pronounced filament eruption corresponding to number 6 in Fig.~\ref{mccdflare}. The maximum of the EW time series corresponds to both absorption signatures, blue and red, i.e. outwards moving and back flowing material (at 20:10~UT). The different colored curves correspond to time series of equivalent widths scaled with factors 40, 50, 100, and 150, which in turn corresponds to fractional active area 24, 30, 60, and 90\%. The black solid line is the original non-scaled equivalent width time series. The grey shaded area corresponds to the $\pm$ 1-$\sigma$ error of the spectrum of HN~Peg. Lower panel: same as upper panel but with half hour averages, which are the averages which have revealed the signature of the erupting filament on EK~Dra. As once can see, the signature of the erupting filament has vanished completely as the half hour average bin includes the times when the second flare and the filament eruption happened.\label{olgmccdarea}}
\end{center} 
\end{figure}
filament eruption, number 3 indicates the peak of the first flare (GOES X3), number 4 indicates the peak of the GOES X-ray light curve, number 5 indicates the onset of the second filament eruption, number 6 indicates the second weaker flare, number 7 indicates the onset of the third and very pronounced filament eruption, number 8 indicates the occurrence of the associated Halo CME and number 9 indicates the first occurrence of back-flowing filament material.\\
As one can see in Fig.~\ref{eventfig}, the flares and the filament eruptions leave distinct signatures in the spatially integrated spectra (first column). A flare causes spectral line core enhancements (see Fig.~\ref{eventfig} third row), an erupting filament causes, when moving towards the observer, a blue-shifted absorption (see Fig.~\ref{eventfig} fourth and sixth row), and back-flowing filament material causes red-shifted absorptions (see Fig.~\ref{eventfig} last row). The strength of the signatures is strongly dependent on the size of the integration area (white and black, respectively, rectangle in the H$\alpha$ core, blue-, and red-wing images shown in columns 2-4). \\
To evaluate if we could detect a typical solar flare and filament eruption in our stellar spectra we use the event from 15/07/2002 in two approaches. For both approaches we calculate the normalized residual flux as f$_{norm}$ = A$_{active}$/A$_{star}\times$(I$_{active}$/I$_{quiet}$ - 1), where A$_{active}$ is the area of the active region, A$_{star}$ is the area of the stellar disk, I$_{active}$ is the spatially integrated solar intensity of the active region, and I$_{quiet}$ is the spatially integrated solar intensity of the quiet Sun (determined from the same region but from the first spectral image of the time series).\\
In the first approach we aim to see with which area scaling we may overcome the 1-$\sigma$ uncertainties of the residual fluxes of the solar event from 15-07-2002 , similarly to the stellar case shown in Fig.\ref{olgevents}. In Fig.\ref{solstell} we plot the residual fluxes obtained from the solar event from 15-07-2002 as 3-minute averages (nominal integration time of the stellar spectroscopic monitoring) scaled to an active region area of 18~\% of the solar disk. With such an area scaling the spectral residual signatures of the eruptive filament exceed 1-$\sigma$. Without any area scaling the residual spectral signatures remain hidden in the noise.\\
As a second approach we superimpose the solar residual event spectrum onto the spectrum obtained for one of our solar analogue target stars and determine the H$\alpha$ equivalent widths. To do so we adapt the wavelength grid of the MCCD observations to the OLG observations and apply the area scaling as described above for approach 1. Here, the term f$_{norm}$ which is the solar residual H$\alpha$ spectrum, is then simply added to the stellar spectrum (see Fig.\ref{olgmccdhnpeg}). Applying the above formalism to the stellar OLG spectra reveals that the solar event from  15/07/2002 would have been invisible if occurring for instance on HN~Peg, simply because the field of view of MCCD corresponds to $\sim$2.3\% and the integration area (white and black, respectively, rectangle in the H$\alpha$ core, blue-, and red-wing images shown in columns 2-4 of Fig.~\ref{eventfig}) only to $\sim$0.6\% of the solar disk, and this is way to small to be seen with respect to the solar disk in a Sun-as-a-star spectrum. Subsequently, we generate in Fig.~\ref{olgmccdarea} a time series of equivalent widths (determined from the full spectral coverage of MCCD, see Fig.~\ref{olgmccdhnpeg}) of the solar residual H$\alpha$ spectra of the 15/07/2002 event superimposed on the spectrum of HN~Peg. The integration area covering  the solar active region is, as mentioned above, too small to make the signature visible in a Sun-as-a-star spectrum. Therefore we scale the active region area by various factors as shown in Fig.~\ref{olgmccdarea}. The grey shaded area denotes the EW error which we estimated using the relation given in \citet{Nissen2018}. As one can see that with a scaling factor of 40 which corresponds to a fractional area of 24\% the filament signature is sufficiently strong to be detected exceeding 1-$\sigma$. The solar H$\alpha$ flare seen as a depression in Fig~\ref{olgmccdarea} is exceeding 1-$\sigma$ with a fractional area of $\sim$12\%, i.e. it is much easier visible than a filament eruption. As we have detected activity on EK~Dra in the OLG data with temporal binning up to 30 minutes we also binned the time series of H$\alpha$ equivalent widths to see how that affects the evolution and to mimic stellar observations with longer exposure times (see lower panel of Fig.~\ref{olgmccdarea}). For a more detailed discussion of this aspect see section~\ref{discussion}.

\section{Discussion}
\label{discussion}
In the discussion, we relate the obtained upper limits of the CME rates of the target stars to results obtained in \citet{Leitzinger2020}, to the expected observable CME rates from \citet{Odert2020}, and to the recent detection of a filament eruption presented in \citet{Namekata2021}. We moreover discuss the solar analysis in comparison with recent similar attempts from literature. \\
\textit{\textbf{Stellar analysis:}} In \citet{Leitzinger2020} we followed a similar approach as here but with archival data which usually have not the desired parameters such as S/N and length of the time series. In \citet{Leitzinger2020} we analyzed a much higher number of stars, among them were also the target stars of the present study. As the analyzed time series in \citet{Leitzinger2020} were shorter and no signatures of stellar CMEs could be detected, the present observations constrain the upper limits of CME rates of the target stars given in \citet{Leitzinger2020}. For HN Peg we find an upper limit for the CME rate of 0.11~day$^{-1}$ which improves the prior limit (2) by a factor of 18, for $\pi^{1}$~UMa we find an upper limit of 0.28~day$^{-1}$ which constrains the prior found limit (2.37) by a factor of $\sim$9, for $\chi^{1}$~Ori we find an upper limit of 0.21~day$^{-1}$ which constrains the prior found limit (4.28) by a factor of $\sim$20, for $\kappa^{1}$~Cet we find an upper limit of 2.2~day$^{-1}$ which constrains the prior found limit (16.42) by a factor of $\sim$8, and finally for EK~Dra we find a detected CME rate of 0.07$^{0.161}_{-0.058}$~day$^{-1}$. These upper limits are to be seen only with respect to the S/N of the spectra, and therefore with CMEs of a certain mass. To relate upper limits of CME rates of stars to the mass of CMEs modelling is necessary. In \citet{Leitzinger2020} we have therefore utilized the semi-empirical model from \citet{Odert2017, Odert2020} to obtain expected observable CME rates which we then compared to the observed ones. The S/N of the spectra determines which CMEs with which masses can be detected. The mean S/N of the observations used in \citet{Leitzinger2020} for the target stars of the present study had a mean of 293, roughly a factor 3 higher than the observations used in the present study. Therefore for the \citet{Leitzinger2020} study CMEs with masses $\ge$ 1.5$\times$10$^{16}$g could have been detected whereas for the present study CMEs with masses $\ge$5$\times$10$^{16}$g could have been detected.\\
The semi-empirical modelling in \citet{Leitzinger2020} was done using the code from \citet{Odert2020} for a fixed parameter set of optical thickness in the H$\alpha$ line $\tau$=10, flare power law index $\alpha$=1.8, and hydrogen column density N$_{h}$=10$^{20}$~cm$^{-2}$. Doing the same for the present observations still the expected observable rates lie above the observed upper limits, meaning that the expected observable rates overestimate the CME rates of those stars. Varying $\tau$ (1 ... 10), $\alpha$ (1.8 ... 2.5), and N$_{h}$ (10$^{20}$ ... 10$^{21}$), yields observable expected CME rates lying above the observed upper limits except the parameter combination N$_{h}$=10$^{21}$~cm$^{-2}$, $\tau$=10. This indicates that plasma with an increased hydrogen column density together with a high optical thickness agrees with the observed upper limits.\\
In contrast to the simple model above, hydrogen column density and optical thickness are not independent and therefore this can be implemented in the calculations for a self-consistent solution with respect to these two quantities. Using fixed values for prominence parameters temperature and gas pressure (P$_{gas}$=0.2~dyn~cm$^{-2}$, T=10000K) and the relationships presented in \citet{Heinzel2015} we self-consistently compute N$_h$ and $\tau$. However, with this modified code the observable expected CME rates overcome the observed upper limits, meaning that with solar parameters the observable expected CME rates are overestimated. We want to note, that already the intrinsic CME rates for active stars are likely overestimated \citep{Odert2017}.\\
Moreover the observable expected CME rate calculations consider a detectability within a 1-$\sigma$ uncertainty, leading to a minimum detectable mass of 5$\times$10$^{16}$g for the OLG observations. Considering a 3-$\sigma$ error range, then the minimum detectable mass increases to 1.5$\times$10$^{17}$g, which is in the order of the most massive solar CMEs. This finding coincides with our expectations that with our observational setup we are able only do detect signatures caused by massive stellar CMEs.\\
Only recently, a filament eruption seen in H$\alpha$ on the solar analog EK~Dra was presented by \citet[][]{Namekata2021}. Coordinated observations with TESS were performed and one TESS flare (BJD(-2450000) 8945.24), was accompanied by a filament eruption. As can be seen from Fig.~\ref{olgevents} we also observed in that period but we did not cover the flare. However, we started observing shortly after this flare and we also detected the signature of an erupting filament, but with lower temporal resolution as \citet[][]{Namekata2021}, as these authors used larger observing facilities. In Fig.~1e of their study they present the residual spectra of the event. The residual spectra tagged with time ranges ``60-80'', ``80-100'' and ``100-120'' are the ones which we have captured also, although we have captured only 10 minutes of their ``60-80'' average. However, when comparing our 30 minute averages with their 20 minute averages, the shape of the absorptions differ to some degree, as the averages are overlapping by 10 minutes. The absorption depth of our first two spectra (see Fig.~\ref{olgevents}) is 0.017 and 0.015 (determined from gaussian fitting), respectively. From the residual time series shown in Fig.~1e in \citet[][]{Namekata2021} the depths are in the order of 0.015. The maximum projected velocities of the absorption reaches $\sim$-600 .. $\sim$500~km~s$^{-1}$ and $\sim$-500 .. $\sim$500~km~s$^{-1}$ for our first two residual spectra, the corresponding residual spectra in Fig.~1e from \citet[][]{Namekata2021} have projected velocity ranges of $\sim$-500 .. $\sim$300~km~s$^{-1}$ and $\sim$-200 .. $\sim$400~km~s$^{-1}$. Our velocities are determined from Gaussian fitting, the ones from \citet[][]{Namekata2021} by eye from the non-fitted residual spectra. Considering the fact that our temporal resolution and S/N is significantly lower than the observations presented in \citet[][]{Namekata2021} the depth and also the projected velocity ranges  agree well. Our observations reveal the second half of the evolution of the eruptive filament structure until the signature vanishes and therefore our observations complement the observations from \citet{Namekata2021} and result in a full coverage of a highly interesting stellar activity event.\\
\citet[][]{Namekata2021} interpreted this event as probable filament eruption following a flare. According to the high projected velocity of the bulk material being close to the escape velocity of the star we support this interpretation. From our observations one can see that the event lasted 1.5 hours longer than presented in \citet[][]{Namekata2021}, simply because their spectral time series ended. Already in their observations the bulk velocity shifted towards the red, meaning that hydrogen plasma is moving towards the star. In our third residual spectrum (see Fig.~\ref{olgevents}) which includes already times when the spectral time series presented in \citet[][]{Namekata2021} had ended, reveals that the projected velocity of the back-falling material stays at the same level as seen in the last residual spectrum in \citet[][]{Namekata2021}. The following residual spectrum in Fig.\ref{olgevents} (no.4) shows then already a reduced projected velocity until in residual spectrum no.5 the absorption has zero velocity. In residual spectrum no.6 the absorption signature is then gone. This behaviour is consistent with the morphology presented in \citet[][]{Namekata2021}.\\
The other flare events shown in the present study (see Fig.~\ref{olgevents1} and Fig.~\ref{olgevents2}) do not show blue wing enhancements or absorptions, i.e. no accompanied filament/prominence eruption is detected with those flares. The morphology of the flares follow a typical flare light curve with a faster rise and a slower decay, except for one event, namely event 04/04/2020. This one reveals no typical H$\alpha$ core enhancements possibly caused by H$\alpha$ footpoint brightenings due to fast electron 
\begin{figure}
\begin{center}
\includegraphics[width=\columnwidth]{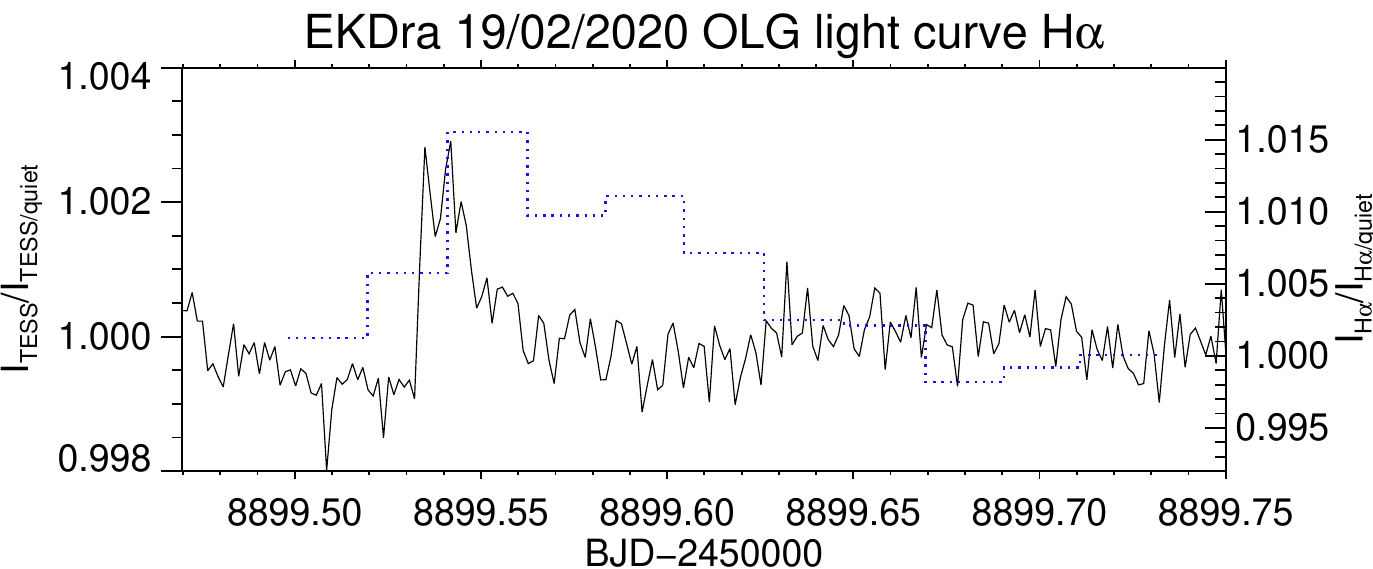}
\includegraphics[width=\columnwidth]{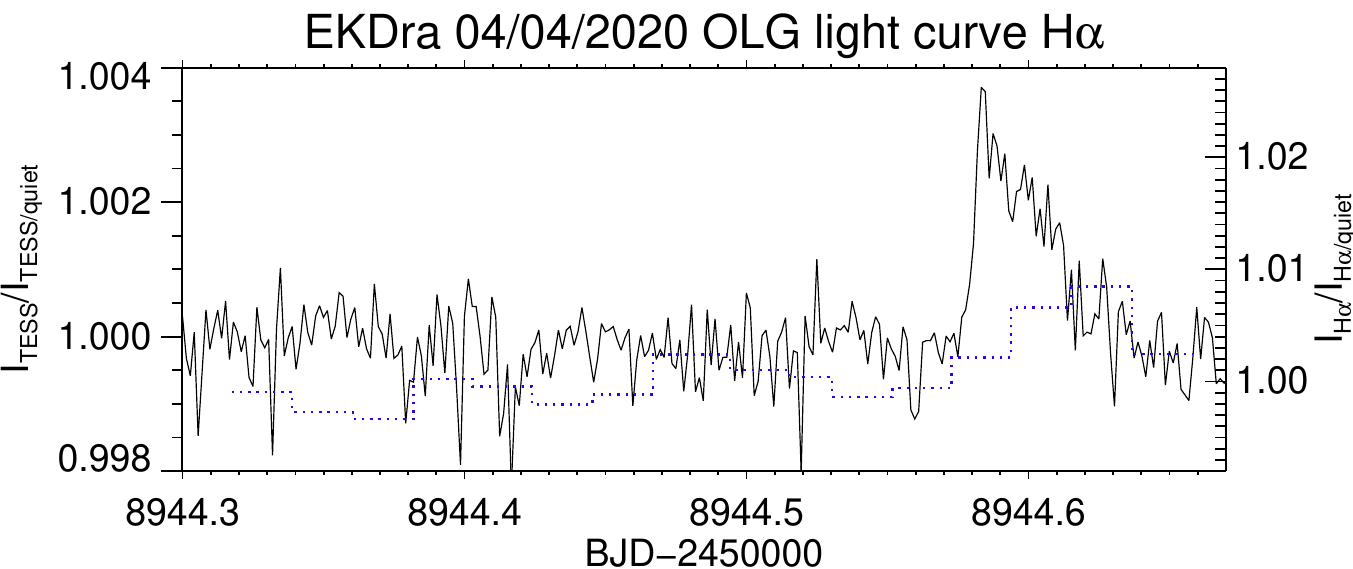}
\includegraphics[width=\columnwidth]{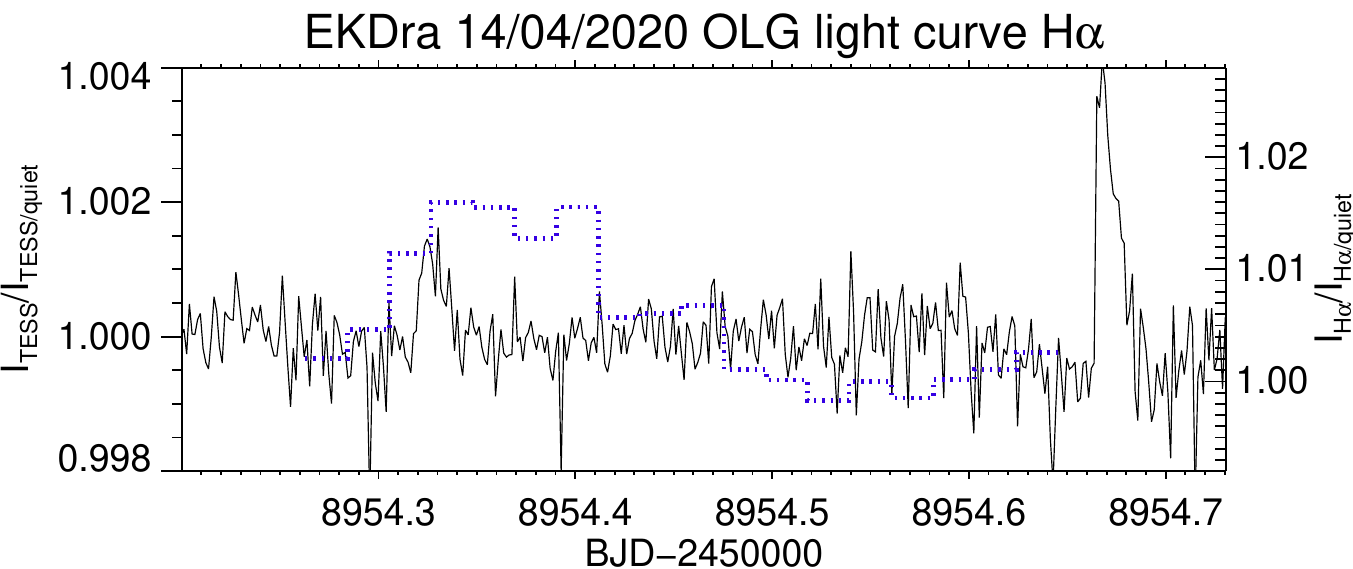}
 \caption{Upper panel: TESS light curve (black solid line) and H$\alpha$ light curve (blue dotted line) extracted from the OLG data for event 19/02/2020. Middle panel: TESS light curve and H$\alpha$ light curve extracted from the OLG data for event 04/04/2020. Lower panel: TESS light curve and H$\alpha$ light curve extracted from the OLG data for 14/04/2020.\label{tessolg}}
\end{center} 
\end{figure}
%
%
beams heating the chromosphere. We only see variation in two spectra, those are spectra no.14 and 15, which show red shifts of 100 and 200~km~s$^{-1}$, respectively. In the following spectrum no extra emission is visible anymore. These red-shifted extra emissions could be caused by coronal rain in cool H$\alpha$ flare loops \citep{Wollmann2023}, especially as these asymmetries are visible in the decaying tail of the flare.\\
A distinct and long-lived (for five residual spectra, i.e. for 2.5~hr) red asymmetry is also visible in event 22/04/2020 (see right panels of Fig.~\ref{olgevents2}). In residual spectrum no. 13  a broad symmetrical emission is visible which slowly shifts to the red with a maximum bulk velocity of 150~km~s$^{-1}$ detected in residual spectrum 17. In residual spectrum no. 18, which is the last of the series, still an enhancement is visible but the residual spectrum is also rather noisy, which does not allow  a distinct identification  of an enhancement.\\
The other two flares, events 04/04/2020 and 14/04/2020 (see Fig.~\ref{olgevents1}), reveal only candidates of minor red asymmetries. We use the term candidate here because the red asymmetries are very subtle and similar subtle emissions are seen as fluctuations in the residual spectra of the series. Event 04/04/2020 is visible in two residual spectra, i.e. 1~hr, whereas event 14/04/2020 is visible in 8 residual spectra which corresponds to 4~hr. The spectral residual peaks are partly shifted by one wavelength bin to the red for the flare events 04/04/2020 and 14/04/2020. The active spectra reveal a core enhancement together with enhanced H$\alpha$ red wings relative to the symmetric quiescent H$\alpha$ profile. There seems to be a plasma component causing the red neighbouring wavelength bin to be somewhat larger than the H$\alpha$ core wavelength bin relative to the quiescent spectrum. Similar red asymmetric spectral residual peaks have been shown only recently in dMe star flare residual spectra by \citet{Notsu2024}, who performed an extensive analysis of a number of dMe star flares spectroscopically. Such red asymmetries have been also modelled by \citet{Wollmann2023} and have been attributed to downward flowing material in flaring loops termed coronal rain. \\
Only very recently \citet{Namekata2024} present again spectroscopic H$\alpha$ observations of EK~Dra, this time with partially coordinated X-ray observations as well as broadband photometry from TESS. The authors find three superflares from which two were accompanied with eruptive prominences, the projected bulk velocities are fast (few hundreds of km~s$^{-1}$) but significantly below the escape velocity of EK~Dra. Together with their previous studies \citep{Namekata2021, Namekata2022a} this results in total to five superflares from which three were accompanied by filament/prominence eruptions. These projected bulk velocities are lower limits, so there is a chance that these filaments/prominences have indeed left the star. The superflares show broadband (TESS) energies ranging from 2 ... 26$\times$10$^{33}$~erg and H$\alpha$ energies ranging from 2 ... 49$\times$10$^{31}$~erg. The authors give an association rate of 60\% between superflares and eruptive filaments/prominences, so three out of five events show possible accompanying eruptive prominences/filaments. We have captured four TESS superflares spectroscopically in H$\alpha$ and one late evolution of a possible filament eruption, so an association rate of 20\% or one out of five events (although we have not observed the superflare related to the possible filament eruption, but we know from TESS that there was one). Our superflares show broadband energies in the range of 2-7$\times$10$^{33}$~erg and H$\alpha$ energies in the range of 7-80$\times$10$^{31}$~erg. We have one superflare (event 14/04/2020) in the sample which shows twice the energy of the most energetic superflare presented in \citet{Namekata2024}, mainly because the duration of the flare lasts for four hours, whereas their most energetic H$\alpha$ flare has a duration of about two hours. In \citet{Namekata2024} the Seimei telescope is used, a 3.8~m telescope which allows observing at a much higher cadence (1-2~min) than we are able with our 0.5m aperture. Therefore we may miss short-lived ($<$30~min) flares and filaments/prominences. The spectral resolving power of the spectrographs is similar. Only one superflare of the sample presented in \citet{Namekata2024} shows no asymmetries, three show blue asymmetries and one shows red asymmetries. The superflares presented in the present study show red asymmetries only.\\
We have captured the time evolution of three TESS flares completely. These are events 19/02/2020, 04/04/2020, and 14/04/2020. For these three flares we overplot the TESS light curves with the OLG H$\alpha$ light curves extracted from the optical spectra. In Fig.~\ref{tessolg} one can see that the flare evolution in both wavelength bands, narrow and broad, is different. The H$\alpha$ light curves suffer from a low cadence, and therefore there is some uncertainty in the peak time of the H$\alpha$ light curves, but the H$\alpha$ flare decay is significantly different from the TESS broadband light curves. The H$\alpha$ decaying flare tails are more extended in time than their TESS counterparts. For event 19/02/2020 and event 14/04/2020 this is likely related to the Neupert effect which is known from solar flares \citep[][]{Neupert1968, Hudson1991}. White light flares are likely caused by fast electron beams accelerated from reconnection regions in solar flares, which hit denser solar atmospheric layers. H$\alpha$ flares are likely caused by electron beams which dissipate their energy in the chromosphere. H$\alpha$ flare emission in solar flares is also a product of post-flare loops occurring in the gradual flare phase. which radiate also in H$\alpha$. In the impulsive flare phase chromospheric evaporation \citep[upward motion of plasma within the flare loops, e.g.][]{Heinzel1994} can occur and also chromospheric condensation can occur \citep[downward motion, e.g.][]{Ichimoto1984, Namekata2022b}.
For white light flares it has been also suggested that not only foot point emission may cause the main emission but that also, especially on young main-sequence stars, the large flaring loops may contribute significantly to white light emission in stellar flares \citep[see][]{Heinzel2018}. 
But there are also flares on stars where white light and H$\alpha$ share a similar evolution. E.g. \citet{Namekata2021} present a TESS white light flare which shows similar temporal evolution as the H$\alpha$ flare. The same applies to the flare which was accompanied by the filament eruption on EK~Dra presented in this study. As \citet[][]{Namekata2021} captured also the flare spectroscopically, they found that the white light and H$\alpha$ evolution of the flare are similar.\\
Event 04/04/2020 differs in the H$\alpha$ light curve from the other two. This event shows no typical H$\alpha$ light curve. Actually we see only two light curve bins of enhanced flux in the middle panel of Fig.~\ref{tessolg}. These occur in the decaying tail of the TESS flare in a form of central weak excess emission (residual spectrum no.14 in the left panel of Fig.~\ref{olgevents2}), followed by a  distinctively red-shifted excess emission. The first broad and central extra emission may be attributed to H$\alpha$ footpoint emission whereas the following red-shifted and stronger excess emission may be attributed to plasma in post-flare loops. As young main-sequence stars may have much larger flare loops as on the Sun, the loop excess emission may be possibly larger than the excess emission originating from the footpoints. \citet{Mullan2006} deduced lengths of flare loops based on EUV/X-ray observations of a number of stars including EK~Dra. These authors analysed two flares on EK~Dra and found loop lengths of 0.28 and 0.42 L/R$_{\star}$ (ratio of loop length to stellar radius), respectively, which is in the range of larger flare loops on the Sun \citep[see e.g.][]{Reale1997, Shibata2011}.\\
We found spectroscopic signatures of flares and a filament eruption on EK~Dra, but why did we find no activity signatures on the other sample stars? The answer lies already in the TESS light curves. One TESS sector has a length of $\sim$ 27~days on average. The total spectroscopic observing times per star are $\sim$14~days for EK~Dra, $\sim$11~days for $\pi^{1}$~UMa, $\sim$14~days for $\chi^{1}$~Ori, $\sim$28~days for HN~Peg, and $\sim$1~day for $\kappa^{1}$~Cet. EK~Dra was so far observed in ten TESS sectors (14, 15, 16, 21, 22, 23, 41, 48, 49, 50) and has revealed flares in every TESS sector; $\pi^{1}$~UMa was observed so far in three TESS sectors (20, 47, 60) and has revealed so far only one flare, $\chi^{1}$~Ori was observed in so far three TESS sectors (43, 44, 45) and has revealed three flares in one TESS sectors, in the remaining two sectors no flares were detected; HN~Peg was so far observed in one TESS sector (55) and revealed no flares; $\kappa^{1}$~Cet was observed so far in two TESS sectors (4,31) and revealed no flares (see Fig.~\ref{TESSlcs} for TESS light curves of the target sample stars). For this comparison between total observing time and number of TESS flares, we set white light flare equal to H$\alpha$ flare, which is not totally correct but a reasonable assumption on the basis of the results presented in \citet{Kretzschmar2011, Watanabe2017}, where a significant fraction ($>$50\%) of the flares show white light components based on solar data sets. If we multiply the TESS flare rate of our stars (see Table~\ref{peakfluxtable}) with their H$\alpha$ observing time then we get the number of flares which should have been detected. For EK~Dra this gives 6, for HN~Peg $<$3, and for the remaining stars 0 flares, which is roughly consistent to what we detected.\\
In Table~\ref{ratetable} we list the expected (see also Table~\ref{peakfluxtable}) H$\alpha$ flare rates exceeding an energy of 10$^{32}$~erg which is a reasonable energy cut-off according to the estimated H$\alpha$ flare energies given in Table~\ref{resulttable}. Only for EK~Dra we can compare expected and observed flare rate and here we see that the expected rate is about an order of magnitude higher than the observed one. Also if we try to relate the observed upper limits of the remaining stars with the expected flare rates is of limited usefulness, as the upper limits are always strongly dependent on the total observing time invested. One can of course derive observed flare and CME rates from very few detected events, as done in this study, but we have to caution that those are of low significance, as those are determined from very few events.
\begin{table}
	\centering%
	\caption{H$\alpha$ flare rates of the present study (3rd column) in comparison to the expected H$\alpha$ flare rates (2nd column). Only EK~Dra has an observed H$\alpha$ flare rate, for the remaining stars we can give only upper limits.}%
	\label{ratetable}
%
%
	
	\begin{tabular}{lccc} 
  name             &  H$\alpha$ flare rate        &     H$\alpha$ flare rate    \\
                   &        expected              &    upper limit/observed     \\
                   & N(>10$^{32}$erg) day$^{-1}$  &      [day$^{-1}$]           \\                       
 \hline                                  
 EK~Dra            &          2.81                &          0.29$^{0.229}_{-0.139}$               \\
 HN~Peg            &          0.48                &          $<$0.11               \\
 $\pi^{1}$~UMa     &          0.46                &          $<$0.21               \\  
$\chi^{1}$~Ori     &          0.36                &          $<$0.28               \\ 
 $\kappa^{1}$~Cet  &          0.23                &          $<$2.22               \\
%
%
 \hline
	\end{tabular}
\end{table}
Another point to consider in the discussion of non detections is that for ground-based observations one will not get uninterrupted data sets such as for satellite observations, due to obvious reasons. This makes then the detection of sporadic events a game of chance. We missed all stronger flares in the EK~Dra TESS light curves (see Fig.~\ref{EKDraTessOLG}), but we still caught some flares because EK~Dra shows several flares in one TESS sector, but e.g. for $\chi^{1}$~Ori or $\pi^{1}$~UMa, which show one or two flares in one sector, then in other sectors no flares, the detection is challenging. So summarizing, especially based on the TESS light curves, its seems reasonable that we detected signatures of activity on EK~Dra and no signatures on HN~Peg although we observed this star twice as much as EK~Dra.\\
\textit{\textbf{Solar analysis:}} The solar data analysis was used to study the detectability of flares and filament eruptions on stars being part of the observed sample, because we did not detect any signature of flares and CMEs on all of the target stars besides EK~Dra. \citet[][]{Namekata2021} used also solar data, in their study from Hida observatory, to base their interpretation of a very probable stellar filament eruption on morphologies of solar filament eruptions. \citet{Otsu2022} investigated a handful of solar activity phenomena (flares, eruptive filaments/surges, and prominences) of the Sun seen as a star, to better interpret stellar observations. Here, we use a complex solar activity event involving flares, filament eruptions, and back flowing material, not to compare its morphology but to evaluate its detectability in OLG spectra.\\
The parameters of the solar event used in the present study are representative for solar H$\alpha$ flares when compared to statistical studies of H$\alpha$ flares on the Sun \citep[see e.g.][]{Temmer2001}. The H$\alpha$ pendant of the X-class flare (see Fig.~\ref{mccdflare}) has a rise time of $\sim$3~min and a duration of $\sim$20-40~min, this agrees well with typical solar H$\alpha$ flare parameters. When comparing this event with the event which occurred on EK~Dra (see Fig.~\ref{olgevents}), we see that the process of backflowing material on the Sun lasts for $\sim$ 40~min whereas on EK~Dra we identify two spectral residuals (no.3, 4) which reveal bulk absorptions with red velocities and therefore a duration of one hour, which is in agreement with the solar event.\\
If we consider stellar integration times (3~min, 30~min), then the solar Sun-as-a-star signature changes its appearance dramatically. This is shown in the lower panel of Fig.~\ref{olgmccdarea} where the same event is shown as in the upper panel, namely the EWs of the solar integrated signatures superimposed on a typical spectrum of HN~Peg obtained at OLG, but with integration times an order of magnitude larger, i.e. around half an hour, which is the same integration time which was needed to make the filament eruption visible on EK~Dra. What one can see clearly is that due to the much lower temporal resolution the signatures falling in one time bin add up and are not well identifiable anymore. The first time bin represents the pre-flare phase. The second time bin includes pre-flare and impulsive flare phase and also a part of the decaying tail. The third time bin then includes the second flare, occurring in the decaying tail of the first flare, and also the major filament eruption including back flowing material. The fourth time bin includes the post-flare phase which interferes already with the next pre-flare phase. The second time bin includes also the first filament eruption which remains completely invisible, but also in the three minute averages (see upper panel of Fig.~\ref{olgmccdarea}), this eruption is not visible as it interferes with the impulsive phase of the flare.\\
We have constructed the solar signature from a fixed number of solar pixels covering the active region from where the activity signatures originate. To enhance the strength of the individual signatures one may have isolated those, preventing thereby that the simultaneously occurring activity signatures (e.g. flare and filament signature) may interfere and thereby affect the observed signature. This is certainly worth to be investigated, as of course flares can occur also without any eruptions. However as we investigated the event on EK~Dra which revealed a flare, a filament eruption and back-flowing material, we were interested in a similar solar event.\\
To evaluate the detectability of activity signatures we have decided to use two approaches. In approach 1 we utilize the same methodology as used for the stellar data, namely to identify signatures of flares and/or erupting filaments/prominences in residual spectra (see Fig.~\ref{solstell}). In approach 2 we determine the EW from the spectra and identify signatures of flares and/or filaments/prominences in the EW time series (see Fig.~\ref{olgmccdarea}). Approach 2 only allows to distinguish between emission and absorption features. With the EW one is not able to distinguish between blue or red absorption, i.e. erupting or back-flowing material. For an estimation on detectability this is sufficient. For the investigation of direction dependent flows one has still to investigate the spectra themselves which we have shown with approach 2. To make the signatures visible in Sun-as-a-star observations we had to scale the solar observations in area. In general, the less pronounced the feature the larger the area ratio scaling needs to be to make the signature visible. Using approach 1 the flare is visible with scaling any area and the filament eruption becomes visible already with an area scaling of 18\%. Using approach 2, already with an area ratio of 12\% the flare peak becomes visible and with an area ratio of 24\% the filament eruption/back falling material becomes visible. Approach one is more sensitive to area ratio scaling than approach 2 and therefore also more useful to be used to search for activity signatures on stars.\\


\section{Summary and Conclusions}
We have presented optical spectroscopic monitoring from OLG of a small sample of young solar analogues. The observations revealed a very low level of detectable activity in H$\alpha$ for the majority of stars in the sample except for EK~Dra, the youngest star in the sample. In the period from January to April 2020 we have detected spectroscopically four flares and one episode of a filament eruption on EK~Dra. The eruptive event has been also published by \citet[][]{Namekata2021} who observed EK~Dra at the same time. We have started observing when the eruptive event was already ongoing and captured the late stages of the eruption and the back falling material until the signature vanished. With our observations from OLG we could complement the erupting filament revealing its full evolution. In the typical exposure times used for spectroscopic monitoring at OLG (three minutes) we did not see the absorption signature caused by an eruptive filament on EK~Dra. Building half hour averages revealed the absorption signatures. To estimate the detectability of solar activity events on our target stars, we spatially integrate over an active solar region revealing a complex event including flares, filament eruptions and back falling material observed by MCCD on MSO, similar to the event on EK~Dra. As expected, when superimposing the spectral spatially integrated solar residuals of a solar event on a typical OLG spectrum of e.g. HN~Peg then it remains invisible. One needs to increase the event area to make the signature visible in a spectrum. We need to scale the active region with factors resulting in a fractional area of the active region being 18\% in residual spectra and 24\% in equivalent width time series, respectively, to make the signature detectable above the noise. This behaviour is consistent with the fact that younger stars have larger active areas and therefore those can be detected in stellar spectra, whereas solar active regions are too small relative to the solar disk to be seen in full disk integrated light i.e. Sun-as-a-star observations. However, even on the other stars of the sample, activity signatures had been expected (from the H$\alpha$ flare rates) but were not visible in the data. We therefore conclude that on solar-like stars already in the first few hundred of Myr the occurrence rates of more massive eruptive filaments/prominences decreases significantly. With our observational setup we might have detected massive events only.\\
The intention of this study was the statistical determination of parameters of stellar eruptive filaments/prominences and their relation to flares. We found four flares and one filament eruption on one star. We know that the filament eruption was accompanied by a flare but the other four flares did not show signatures of filament eruptions. So one out of five flares on EK~Dra shows an accompanying filament eruption, but this result is far from being statistically significant. Although the observational efforts have been increased in the past few years to detect stellar CMEs still the number of distinct events is low. We know many more candidate events, at least for the method of Doppler-shifted absorption/emission \citep[e.g.][]{Fuhrmeister2018, Vida2019}. One way to obtain statistics is to focus on the numerous (few hundreds) candidate events and try to better understand those. This has already partly begun with the systematic investigation of spatially integrated solar, i.e. Sun-as-a-star signatures of flares and eruptive filaments, with the aim to better characterize stellar signatures of flares and eruptive filaments/prominences, including their temporal evolution \citep[see e.g.][]{Leitzinger2021, Namekata2021, Leitzinger2022b, Otsu2022}. These studies used solar instruments capable of spatially resolved 2D spectroscopy (such as MCCD on Mees Solar Observatory/MSO) or full-disk photometry in various filters (Solar Dynamics Doppler Imager/SDDI on The Solar Magnetic Activity Research Telescope/SMART).\\
With this study we have demonstrated that small-sized telescopes can be used to infer spectroscopic activity signatures on bright solar-like stars. 
From the spectroscopic monitoring presented in this study we have seen that CMEs, more energetic and massive than occurring on our present-day Sun, on few hundred Myr old solar analogues are not a frequent phenomenon.\\

\section*{Acknowledgements}
We thank the anonymous referee for very valuable comments which improved the manuscript. This research was funded in whole, or in part, by the Austrian Science Fund (FWF) [10.55776/P30949, 10.55776/I5711]. For the purpose of open access, the author has applied a CC BY public copyright licence to any Author Accepted Manuscript version arising from this submission. This paper includes data collected by the TESS mission. Funding for the TESS mission is provided by the NASA's Science Mission Directorate. Mees acknowledgements go to NASA HDEE Grants 80NSSC18K0064 and 80NSSC18K1658 for the data rescue effort, NASA Grant NAGW 1542 for the instrument fabrication. Support was also provided from Lockheed under NASA contract NAS8-37334 with Marshall Space Flight Center and the Yohkoh mission contract NAS8-40801. This work presents results from the European Space Agency (ESA) space mission Gaia. Gaia data are being processed by the Gaia Data Processing and Analysis Consortium (DPAC). Funding for the DPAC is provided by national institutions, in particular the institutions participating in the Gaia MultiLateral Agreement (MLA). The Gaia mission website is \url{https://www.cosmos.esa.int/gaia}. The Gaia archive website is \url{https://archives.esac.esa.int/gaia}. 


\section*{Data Availability}
For the present study we used data from TESS, Mees, GAIA DR3, and OLG. TESS data are freely available via the Mikulski Archive for Space Telescopes (MAST, \url{https://mast.stsci.edu/portal/Mashup/Clients/Mast/Portal.html}). Mees data are freely avalailable and accessible under \url{https://www.cora.nwra.com/MCCD/}. Gaia data are freely available and accessible under \url{https://archives.esac.esa.int/gaia}. OLG data are not freely available but can be made available under request via email.
 



\bibliographystyle{mnras}
\bibliography{Mybibfile} 





\section{Appendix}

\begin{figure*}
\begin{center}
	\includegraphics[width=15cm]{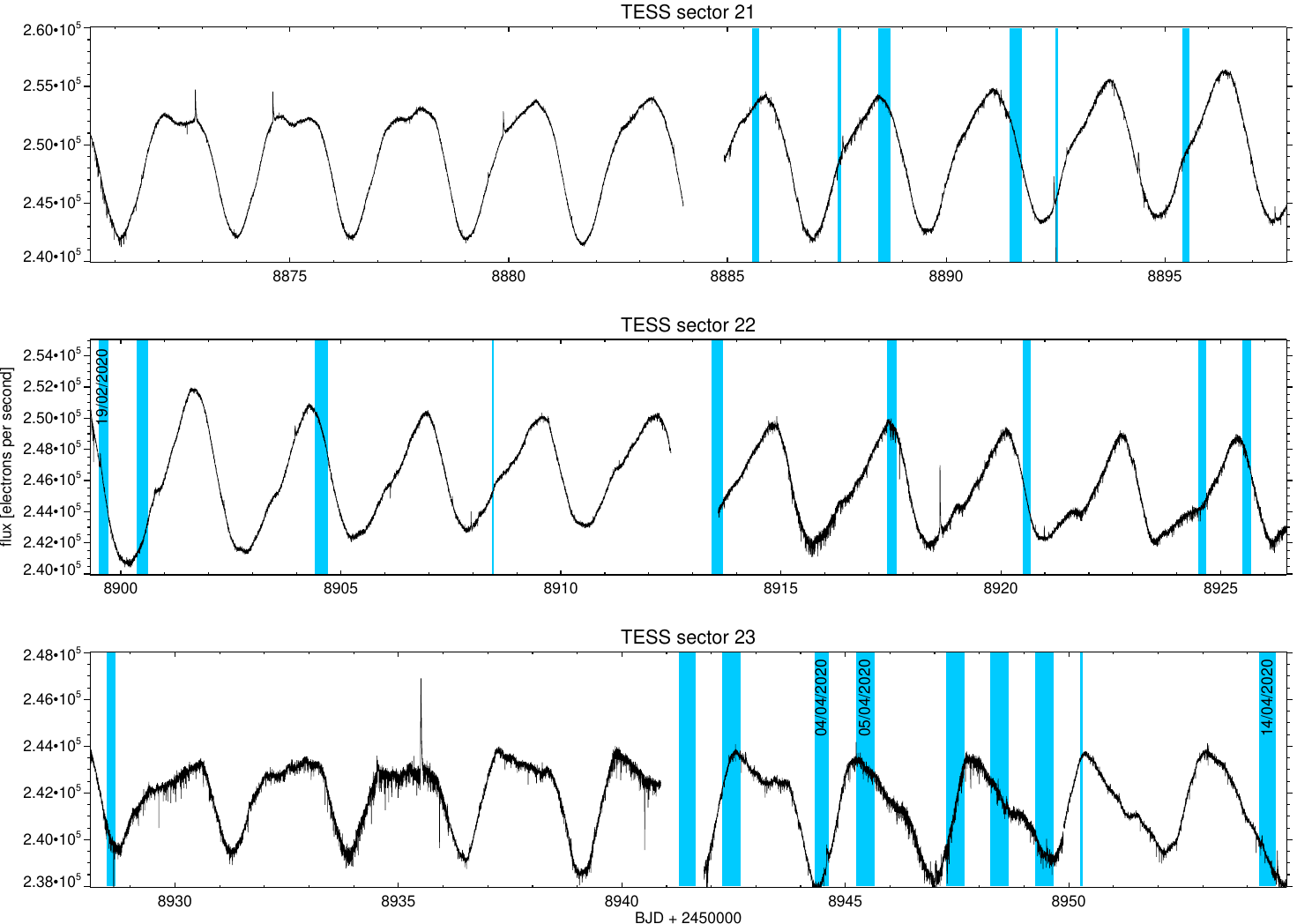}
  	\caption{TESS light curves of EK Dra during which OLG spectroscopic observations are available. As one can see we missed all larger flares. Only few weaker flares (although all of them are superflares) have been captured. The numbers given indicate the spectroscopic data sets per month for the detected events so that the reader can relate the TESS light curve with the residual spectra in Fig.~\ref{olgevents},~\ref{tessolg},~\ref{olgevents1},~\ref{olgevents2}. \label{EKDraTessOLG}
}
\end{center}
\end{figure*}



\begin{figure*}
  \centering
    \includegraphics[width=\columnwidth]{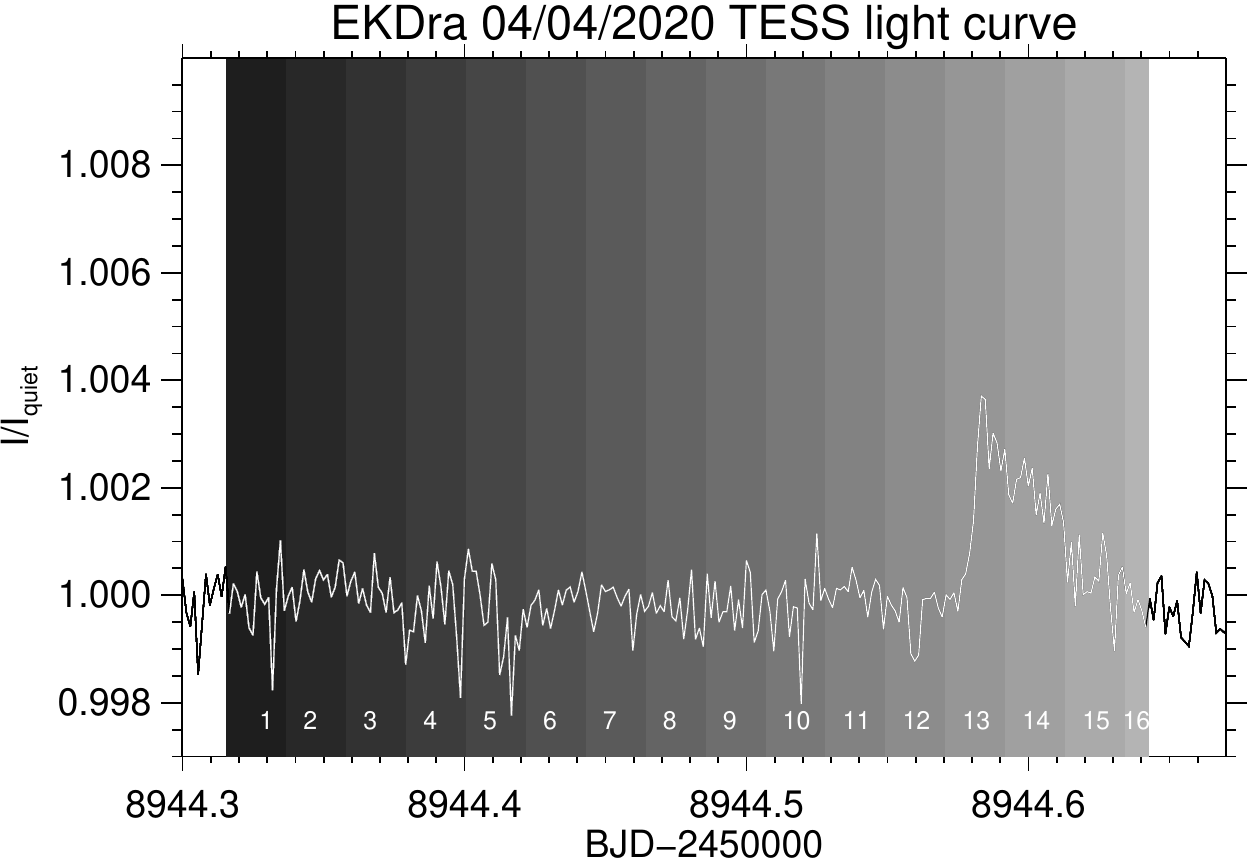} \hfill
    \includegraphics[width=\columnwidth]{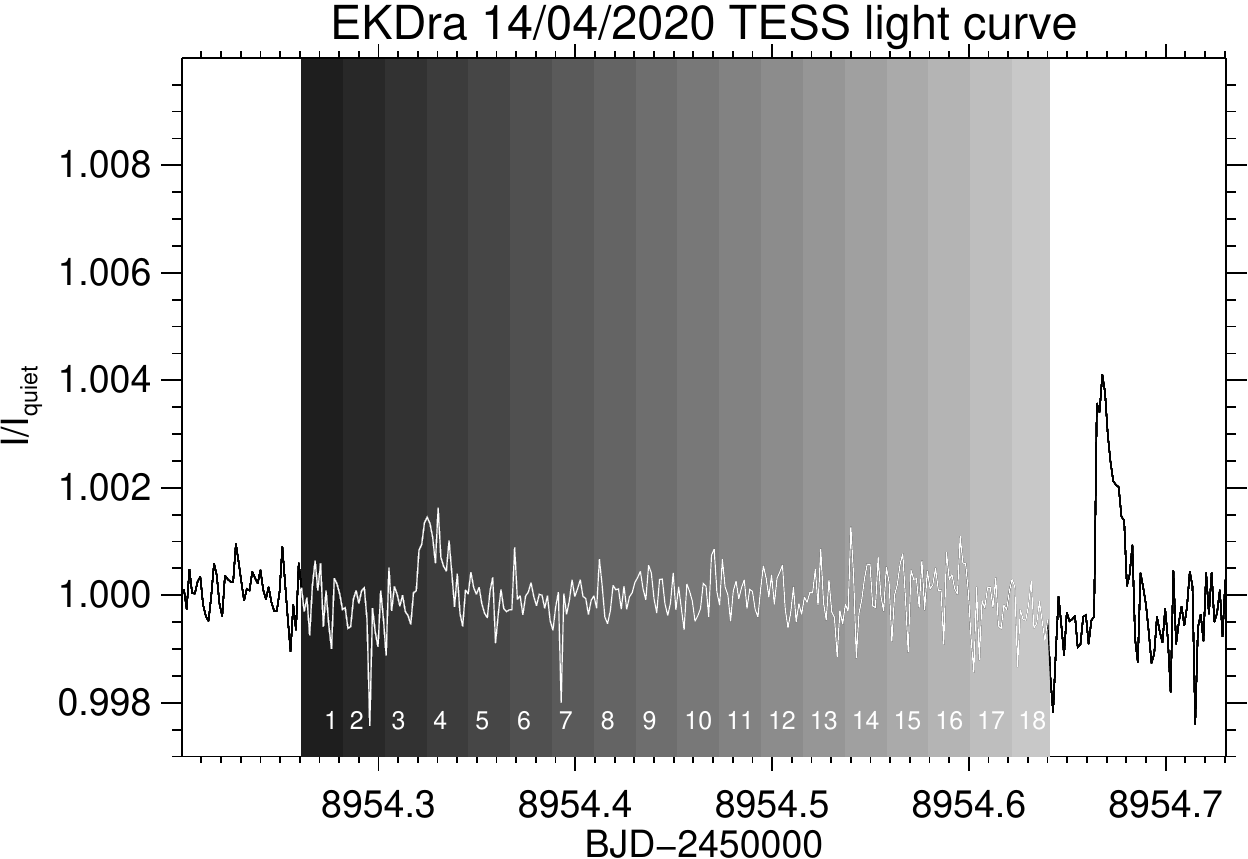} 
      \\[\smallskipamount]
    \includegraphics[width=\columnwidth]{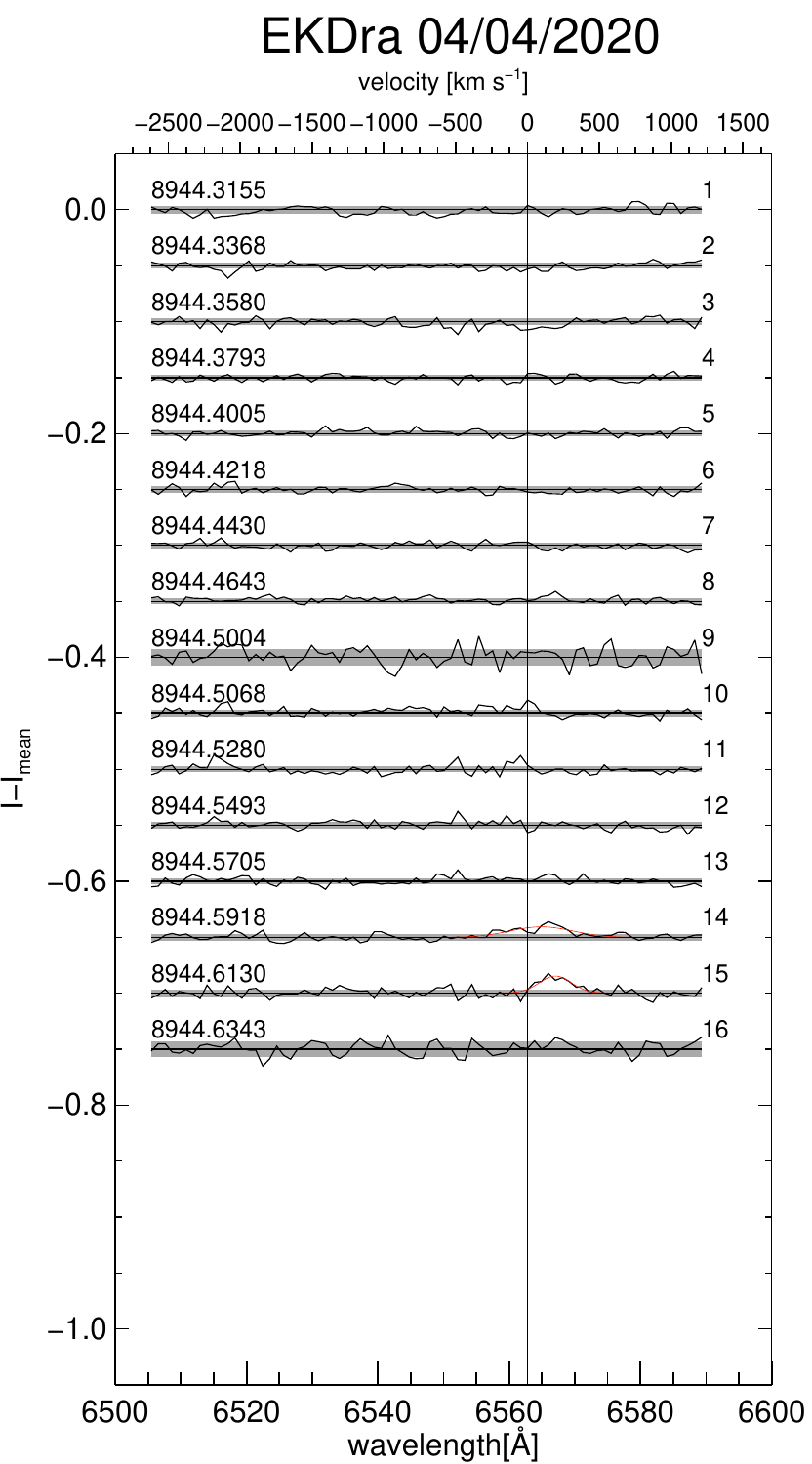} 
    \includegraphics[width=\columnwidth]{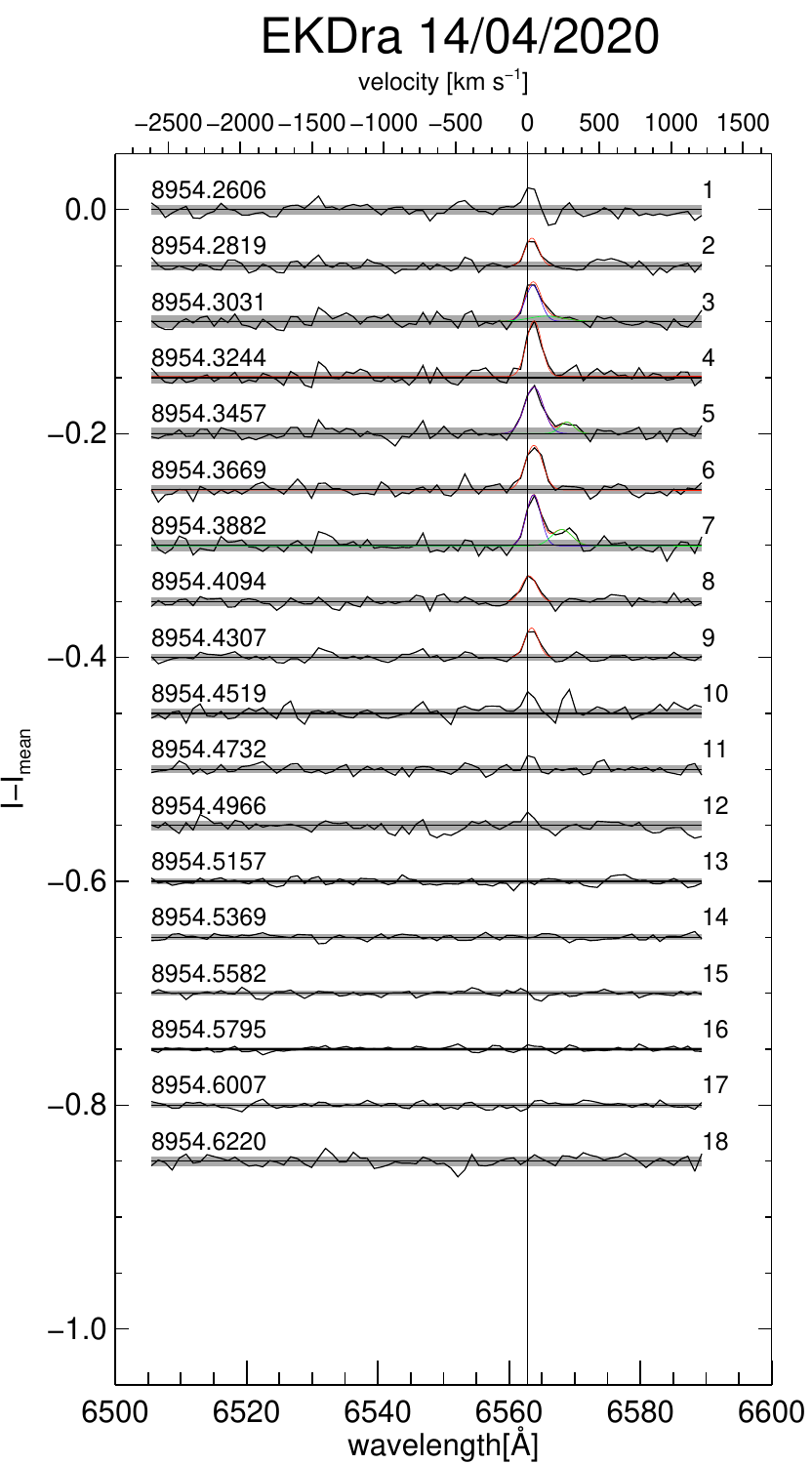} 
  \caption{Photometric (upper panel) and residual spectral time series (lower panel; from top to bottom) of EK~Dra from the night of the 4th of April 2020 (left panels) and of EK~Dra from the night of the 14th of April 2020. Grey shaded areas denote the 1-$\sigma$ uncertainty of the residuals. Red solid lines are gaussian fits (single and double) to the excess emission visible from residuals 14-15 (lower left panel) and residuals 2-9 (lower right panel). Solid blue and green lines are the single components of double gaussian fitting.}
  \label{olgevents1}
\end{figure*}


\begin{figure*}
  \centering
    \includegraphics[width=\columnwidth]{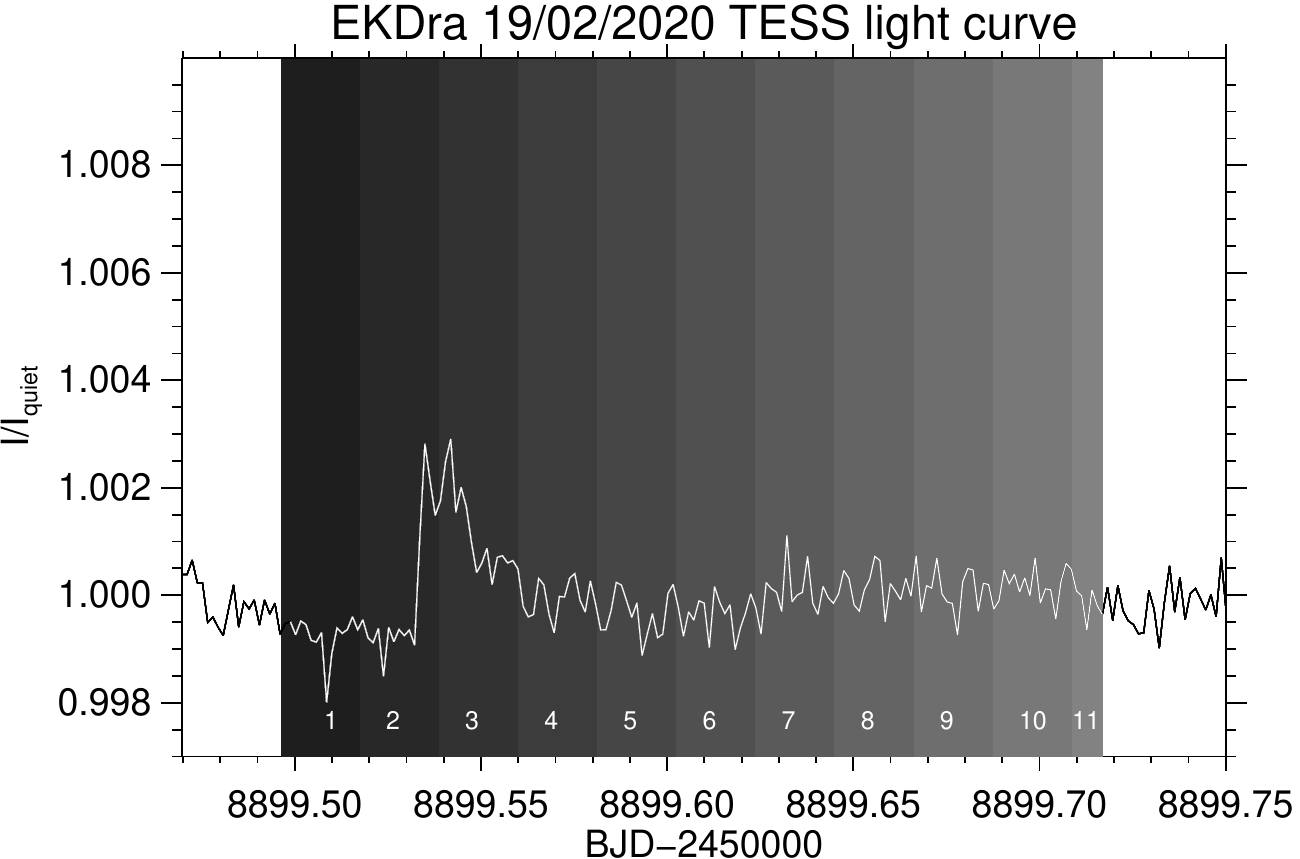}
      \\[\smallskipamount]
    \includegraphics[width=\columnwidth]{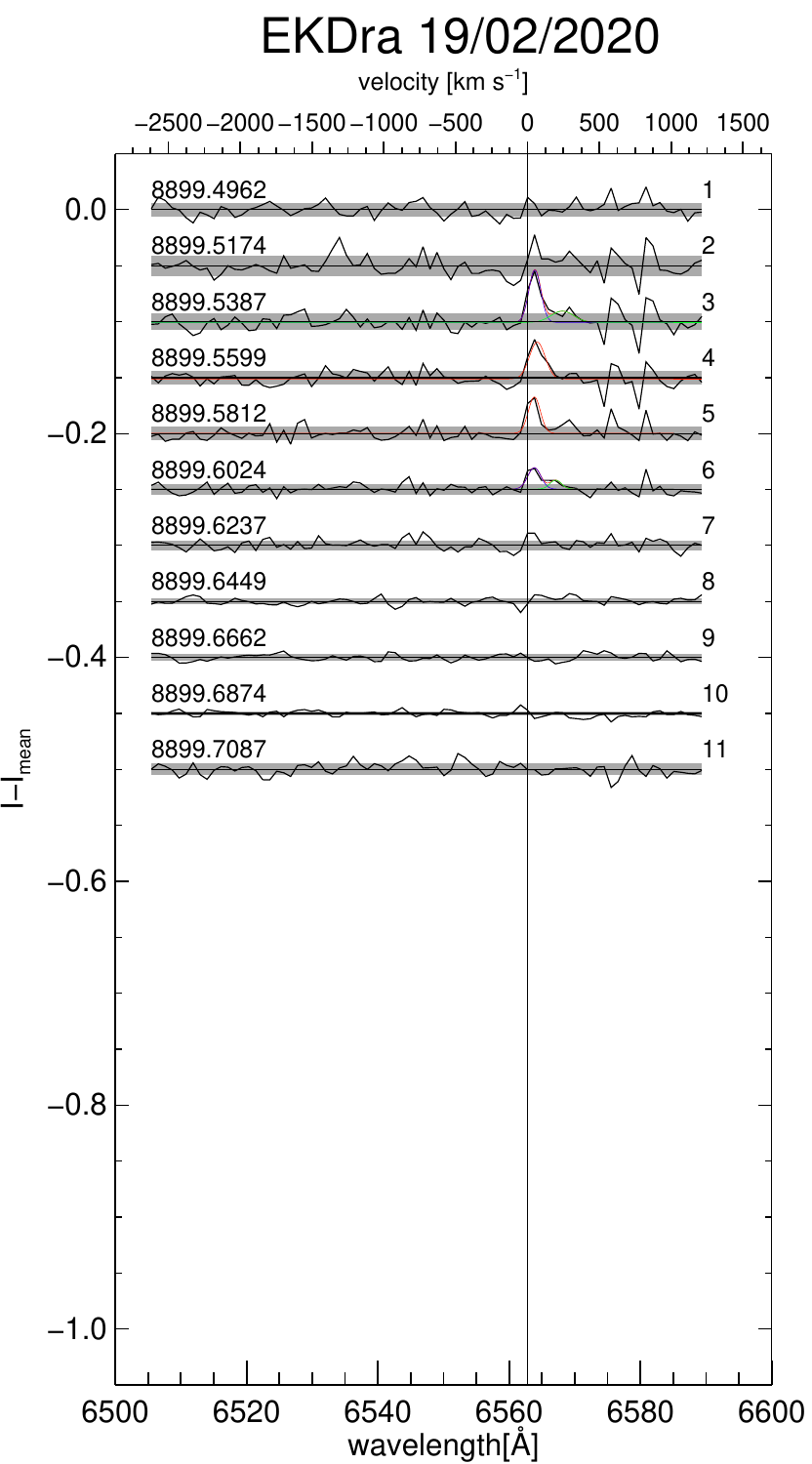}
    \includegraphics[width=\columnwidth]{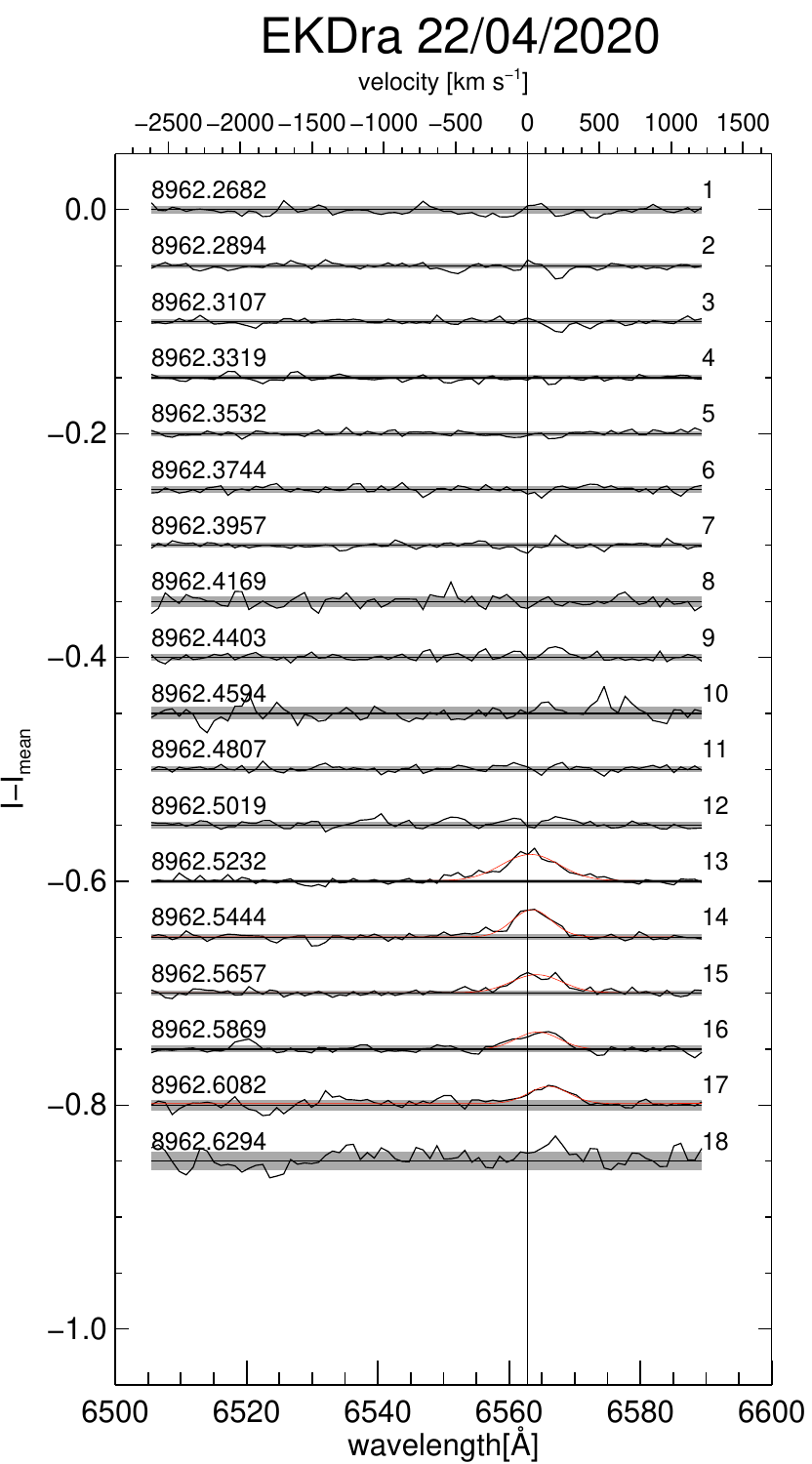} 
  \caption{Photometric (upper panel) and residual spectral time series (lower panel; from top to bottom) of EK~Dra from the night of the 19th of February 2020 (left panels) and of EK~Dra from the night of the 22nd of April 2020. Grey shaded areas denote the 1-$\sigma$ uncertainty of the residuals. Red solid lines are gaussian fits (single and double) to the excess emission visible from residuals 3-6 (lower left panel) and residuals 13-17 (lower right panel). Solid blue and green lines are the single components of double gaussian fitting.}
  \label{olgevents2}
\end{figure*}

\bsp	
\label{lastpage}
\end{document}